\documentclass[prd,tightenlines,nofootinbib,superscriptaddress,showpacs]{revtex4}
\usepackage[utf8x]{inputenc}
\usepackage{amsmath} 
\usepackage{amsfonts,hyperref}
\usepackage{slashed}
\usepackage[dvips]{graphicx}
\usepackage{color}
\usepackage{units}
\usepackage{bbm}
\usepackage{soul}
\allowdisplaybreaks

\renewcommand{\O}{{\mathcal{O}}}
\renewcommand{\Im}{{\text{Im}}}

\newcommand{\e}{{\text{e}}}
\newcommand{\ii}{\text{i}}
\renewcommand{\e}{{\text{e}}}

\DeclareMathOperator\sgn{sgn}

\usepackage{color}

\begin{document}
 \author{{\bf Clément Stahl}\email{clement.stahl@icranet.org}}
 \affiliation{ICRANet, Piazzale della Repubblica 10, 65122 Pescara, Italy}
 \affiliation{Dipartimento di Fisica, Universit\`a di Roma "La Sapienza", Piazzale Aldo Moro 5, 00185 Rome, Italy}
 \affiliation{Universit\'e de Nice Sophia Antipolis, 28 Avenue de Valrose, 06103 Nice Cedex 2, France}
 \author{{\bf Eckhard Strobel}\email{eckhard.strobel@gravity.fau.de}}
 \affiliation{ICRANet, Piazzale della Repubblica 10, 65122 Pescara, Italy}
 \affiliation{Dipartimento di Fisica, Universit\`a di Roma "La Sapienza", Piazzale Aldo Moro 5, 00185 Rome, Italy}
 \affiliation{Universit\'e de Nice Sophia Antipolis, 28 Avenue de Valrose, 06103 Nice Cedex 2, France}
 \author{{\bf She-Sheng Xue}\email{xue@icra.it}}
 \affiliation{ICRANet, Piazzale della Repubblica 10, 65122 Pescara, Italy}
 \affiliation{Dipartimento di Fisica, Universit\`a di Roma "La Sapienza", Piazzale Aldo Moro 5, 00185 Rome, Italy}

\date{\today}

\title{Fermionic current and Schwinger effect in de Sitter spacetime}

\begin{abstract}
 We study the fermionic Schwinger effect in two dimensional de Sitter spacetime. To do so we first present a method to semiclassically compute the number of pairs created per momentum mode for general time dependent fields. In addition the constant electric field is studied in depth. In this case, solutions for the Dirac equation can be found and the number of pairs can be computed using the standard Bogoliubov method. This result is shown to agree with the semiclassical one in the appropriate limit. The solutions are also used to compute the expectation value of the induced current.  Comparing these results to similar studies for bosons we find that while the results agree in the semiclassical limit they do not generally. Especially there is no occurrence of a strong current for small electric fields.
\end{abstract}

 %Semiclassical theories and applications, Quantum fields in curved spacetime, Field theories in dimensions other than four,  Particle-theory and field-theory models of the early Universe (including cosmic pancakes, cosmic strings, chaotic phenomena, inflationary universe, etc.)

\maketitle

%%%%%%%%%%%%%%%%%%%%%%%%%%%%%%%%%%%%%%%%%%%%%%%%%%%%%%%%%%%%%%%%%%%%%%%%%%%%%%%%%%%%%%%%%%%%%%%%%%%%%%%%%%%%%%%%%%%%%%%%%%%%%%%%%%%%
\section*{Introduction}
%%%%%%%%%%%%%%%%%%%%%%%%%%%%%%%%%%%%%%%%%%%%%%%%%%%%%%%%%%%%%%%%%%%%%%%%%%%%%%%%%%%%%%%%%%%%%%%%%%%%%%%%%%%%%%%%%%%%%%%%%%%%%%%%%%%%

The Schwinger effect, i.e.~pair production by a strong electric field, was first studied by Sauter \cite{Sauter1931}. It has been of interest for research over the years and is still today (see e.g.~\cite{Heisenberg1936,Schwinger1951,Brezin1970,Popov1971,Popov1972,Popov1973,Marinov1977,Popov2001,Dunne2005B,Dunne2006,Kleinert2008,Kleinert2013,Blinne2013}). While these studies are performed in flat Minkowski spacetime the development of quantum field theory in curved spacetimes, as a way to combine quantum field theory with Einstein's theory of gravitation  \cite{RedBook,Birrell1984}, has opened new possibilities. It has been found that the gravitational field gives rise to a pair production in a similar way as the electric field \cite{Parker1968,Parker1969,Parker1971,DeWitt1975}. This effect is used as an explanation for the generation of primordial cosmic inhomogeneities which in turn can explain the large scale structure of the universe. Recently also the combination of the effects of the gravitational and electric field have been studied in various frameworks \cite{Garriga1994,Garriga1994B,Villalba1995,Haouat2013,Froeb2014,Kobayashi2014,Haouat2015,Habib1999}.\\
As the flat  Minkowski spacetime, \(D\)-dimensional de Sitter (\(\text{dS}_D\)) spacetime has constant scalar curvature and is maximally symmetric.  It is used to describe the early stages of inflation (see for \cite{Martin2007,Baumann2014}  reviews) as well as the late stage of acceleration of the expansion (see \cite{Carroll2001}) in cosmology. In addition its study might shed light on the understanding of the quantum nature of spacetime \cite{Bunch1978,Mottola1985} (see also \cite{Birrell1984} page 340ff.).\\
The study of the Schwinger effect in dS spacetime is of interest for various reasons. It can be used to study false vacuum decay and bubble nucleation \cite{Garriga1994,Froeb2014} as well as to put constraints on magnetogenesis \cite{Kobayashi2014}. It is also used to test the ER=EPR conjecture using the AdS/CFT correspondence \cite{Fischler2014}. In addition it can help to understand the connection between several methods of regularization better \cite{Landete2014}. Very recently a unified thermal picture of Schwinger effect in both dS and anti-de Sitter spacetime and Hawking radiation near Reissner-Nordström black holes has been proposed \cite{kim2015}. \\
The regularized current of scalar QED has recently been computed in \(\text{dS}_2\) using Pauli-Villars subtraction in \cite{Froeb2014} and in \(\text{dS}_4\) using adiabatic regularization in \cite{Kobayashi2014}. While fermionic pair creation in \(\text{dS}_2\) has been studied \cite{Villalba1995,Haouat2013,Haouat2015} the computation of the current has not been performed. Here we will compute the fermionic current in \(\text{dS}_2\) and regularize it using adiabatic regularization.\\
This paper is organized as follows. In section \ref{sec:preliminaries} we derive the equations necessary to study the Schwinger effect in \(\text{dS}_2\) spacetime. We then present a method to compute the pair creation rate semiclassically for general electric fields in \(\text{dS}_2\) spacetime in section \ref{sec:SC}. After this we concentrate on the constant electric field and compute the pair creation rate as well as the expectation value of the induced current in section \ref{sec:constantfield}.  We conclude in section \ref{sec:conclusions}. To make the main ideas clearer we relegated some of the technical calculations to the appendices.  

%%%%%%%%%%%%%%%%%%%%%%%%%%%%%%%%%%%%%%%%%%%%%%%%%%%%%%%%%%%%%%%%%%%%%%%%%%%%%%%%%%%%%%%%%%%%%%%%%%%%%%%%%%%%%%%%%%%%%%%%%%%%%%%%%%%%
\section{The Dirac equation in \texorpdfstring{\(\text{dS}\)}{dS} spacetime}
%%%%%%%%%%%%%%%%%%%%%%%%%%%%%%%%%%%%%%%%%%%%%%%%%%%%%%%%%%%%%%%%%%%%%%%%%%%%%%%%%%%%%%%%%%%%%%%%%%%%%%%%%%%%%%%%%%%%%%%%%%%%%%%%%%%%
\label{sec:preliminaries}
In this section we define the framework in which we will work in this paper. To do so we review the basic equations of QED in curved spacetime in \ref{sec:QEDiCST}. Subsequently we will derive the Dirac equation in \(\text{dS}_2\) in section \ref{sec:DiracEq}. For more information on QFT in curved spactime see \cite{RedBook,Birrell1984}.
%%%%%%%%%%%%%%%%%%%%%%%%%%%%%%%%%%%%%%%%%%%%%%%%%%%%%%%%%%%%%%%%%%%%%%%%%%%%%%%%%%%%%%%%%%%%%%%%%%%%%%%%%%%%%%%%%%%%%%%%%%%%%%%%%%%%
\subsection{QED in curved spacetime}
%%%%%%%%%%%%%%%%%%%%%%%%%%%%%%%%%%%%%%%%%%%%%%%%%%%%%%%%%%%%%%%%%%%%%%%%%%%%%%%%%%%%%%%%%%%%%%%%%%%%%%%%%%%%%%%%%%%%%%%%%%%%%%%%%%%%
\label{sec:QEDiCST}
We start from the action of QED in curved spacetime coupled to a spinor field \(\psi(x)\)
\begin{align}
\label{eq:action}
S=\int \text{d}^2 x  \sqrt{-g(x)} \left[  -\frac{1}{\kappa}R(x) +\frac{\ii}{2} \left[\overline{\psi}(x) \underline{\gamma}^{\mu} \nabla_{\mu} \psi(x) -\left(\nabla_{\mu} \overline{\psi}(x)\right)\underline{\gamma}^{\mu}\psi(x)\right] -m \overline{\psi}(x) \psi(x)  -\frac{1}{4} F_{\mu \nu}(x)F^{\mu \nu}(x) \right].
\end{align} 
The scalar density of weight 1/2 which generalizes the flat spacetime Lagrangian is $S:=\int \text{d}^2 x \mathcal{L}$. For the computation of the Schwinger effect we assume a background gravitational field defined through the metric \(g_{\mu\nu}\) as well as a background electric field given by
\begin{align}
F_{\mu \nu}(x):=\partial_{\mu} A_{\nu}(x)-\partial_{\nu} A_{\mu}(x).
\end{align}
We also introduce the tetrad field \(e^a_{\mu}(x)\) with the help of the two dimensional Minkowski metric \(\eta_{ab}\) 
\begin{align}
g_{\mu \nu}(x)=e^a_{\mu}(x) e^b_{\nu}(x) \eta_{ab}. \label{eq:tetraddefinition}
\end{align}
The covariant derivative for fermions is given by
\begin{align}
  \nabla_{\mu} := \hbar\left(\partial_{\mu}-\frac{\ii}{4}\omega^{ab}_\mu\sigma_{ab}\right) +\ii e A_{\mu}(x) ,
\end{align}
where we defined the commutator of the gamma matrices 
\begin{align}
 \sigma_{ab}&:=\frac{\ii}{2}[\gamma_a,\gamma_b]
\end{align}
as well as the spin connection
\begin{align}
\begin{split}
\omega_{\mu}^{ab}&:= \frac{1}{4}\left[e^{b\alpha}(x) \partial_{\mu}e^a_{\alpha}(x)-e^{a\alpha}(x) \partial_{\mu}e^b_{\alpha}(x)+ e^{a \alpha}(x) \partial_{\alpha}e^b_{\mu}(x)-e^{b\alpha}(x) \partial_{\alpha} e^{a \mu}(x)\right.\\ & \hspace{2cm}\left.+e^{b\nu}(x)e^{a \lambda}(x)e_{c \mu}(x)\partial_{\lambda} e^c_{\nu}(x)-e^{a\nu}(x)e^{b \lambda}(x)e_{c \mu}(x)\partial_{\lambda} e^c_{\nu}(x) \right]. \label{eq:spinconnection}
\end{split}
\end{align}
The gamma matrices in curved spacetime \(\underline{\gamma}^\mu\) are related to the usual ones in Minkowski spacetime via 
 \begin{align}
\gamma^a =: \underline{\gamma}^{\mu} e^a_{\mu}. 
\end{align} 
The canonical momentum is given by 
\begin{align}
 \pi(x):=\frac{\partial \mathcal{L}}{\partial(\partial_0\psi(x))}=\sqrt{-g(x)} \,\ii\hbar\overline{\psi}(x)\underline{\gamma}^0. \label{eq:mometum}
\end{align}
In curved spacetime, the Hermitian adjoint is given by \cite{RedBook,Parker1980,Pollock2010}\footnote{This definition using the gamma matrix \(\gamma^0\) of flat spacetime ensures that $ \overline{\psi}(x) \psi(x)$ transforms as a scalar, is real and that the probability current $j^{\mu}= \overline{\psi}(x) \underline{\gamma}^{\mu} \psi(x)$ is conserved.}
\begin{align}
 \overline{\psi}(x)=\psi^\dagger(x)\gamma^0.
\end{align}
The Dirac equation can be derived from the action (\ref{eq:action}) by varying with respect to the field \(\overline{\psi}(x)\), which gives
 \begin{align}
\left(i\underline{\gamma}^{\mu} \nabla_{\mu} -m \right) \psi(x) =0. \label{eq:CurvedDiracGen}
 \end{align}
%%%%%%%%%%%%%%%%%%%%%%%%%%%%%%%%%%%%%%%%%%%%%%%%%%%%%%%%%%%%%%%%%%%%%%%%%%%%%%%%%%%%%%%%%%%%%%%%%%%%%%%%%%%%%%%%%%%%%%%%%%%%%%%%%%%%
 \subsection{The Dirac equation in \texorpdfstring{\(\text{dS}_2\)}{dS2}}
%%%%%%%%%%%%%%%%%%%%%%%%%%%%%%%%%%%%%%%%%%%%%%%%%%%%%%%%%%%%%%%%%%%%%%%%%%%%%%%%%%%%%%%%%%%%%%%%%%%%%%%%%%%%%%%%%%%%%%%%%%%%%%%%%%%%
 \label{sec:DiracEq}
Two dimensional dS spacetime is described by 
\begin{align} 
\text{ds}^2=a(\eta)^2 (\text{d}\eta^2-\text{d}x_1^2), \label{eq:dSmetric}
\end{align}
where the scale factor \(a(\eta)\) depends on the conformal time \(\eta\) and the constant Hubble factor \(H\) following
\begin{align}
a(\eta):= -\frac{1}{H \eta },\hspace{1cm}  \hspace{1cm} (-\infty<\eta<0).
\end{align}
Following from (\ref{eq:tetraddefinition}) and (\ref{eq:dSmetric}) the tetrads for dS space are given by 
\begin{align}
 e^a_{\mu}(x)=a(\eta)\delta_{\mu}^a. \label{eq:tetrad}
\end{align}
The non-zero components of the spin connection (\ref{eq:spinconnection}) are found to be 
  \begin{align}
 \label{eq:connection}
 \omega_{1}^{01}(x)=-\omega_{1}^{10}(x)=\frac{a'(\eta)}{2a(\eta)}, 
 \end{align}
where prime denotes derivative with respect to \(\eta\).\\
Using the explicit form of the tetrads (\ref{eq:tetrad}) and the spin connection (\ref{eq:connection}) the Dirac equation of \(\text{dS}_2\) spacetime can be derived from
(\ref{eq:CurvedDiracGen}) as
  \begin{equation}
\left[\ii \left(\hbar\underline{\gamma}^{\mu} \partial_{\mu}+\frac{\hbar}{2} \frac{a'(\eta)}{a(\eta)} \underline{\gamma}^0 + \ii e A_{\mu}(x) \underline{\gamma}^{\mu} \right) -m \right] \psi(x)=0. \label{eq:CurvedDirac}
 \end{equation}
We choose to work in the Weyl basis
\begin{align}
 \gamma^0=\begin{pmatrix}
           0&1\\
           1&0
          \end{pmatrix},&&
 \gamma^{1}=\begin{pmatrix}
           0&1\\
           -1&0
          \end{pmatrix}.
          \end{align}
Using a momentum mode decomposition of the form\footnote{The exact form of the momentum decomposition is specified later in (\ref{eq:decomp2}).}
\begin{align}
\psi(x)\sim\e^{\frac{\ii}{\hbar} k x_1}  \begin{pmatrix}
           \psi_1(\eta)\\
           \psi_2(\eta) 
          \end{pmatrix} \label{eq:decomp1}
\end{align}
as well as a solely time dependent electric field \(A_\mu(x)=(0,A(\eta))\) we find that (\ref{eq:CurvedDirac}) takes the form
\begin{align}
 \ii\hbar {\psi_1}'(\eta)+p(\eta)\psi_1(\eta)+\frac{\hbar}{2} \frac{a'(\eta)}{a(\eta)}\psi_1(\eta)-m a(\eta)\psi_2(\eta)=&0,\label{eq:CoupledDirac1}\\
 \ii\hbar {\psi_2}'(\eta)-p(\eta)\psi_2(\eta)+\frac{\hbar}{2} \frac{a'(\eta)}{a(\eta)}\psi_2(\eta)-m a(\eta)\psi_1(\eta)=&0 \label{eq:CoupledDirac2},
\end{align}
where
\begin{align}
 p(\eta):=k+eA(\eta)
\end{align}
is the kinetical momentum. 
%\subsection{Decoupling the Dirac equation}
Decoupling these equations leads to
\begin{align}
 \hbar^2\psi_1''(\eta)+\left(\omega_k(\eta)^2-\ii\hbar\, p(\eta) \left[\frac{p'(\eta)}{p(\eta)}-\frac{a'(\eta)}{a(\eta)}\right]+\hbar^2\left[\frac{a''(\eta)}{2a(\eta)}-\frac{3 a'(\eta)^2}{4 a(\eta)^2}\right]\right)\psi_1(\eta)=&0, \label{eq:DecoupledDirac1}\\
 \hbar^2\psi_2''(\eta)+\left(\omega_k(\eta)^2+\ii\hbar\, p(\eta) \left[\frac{p'(\eta)}{p(\eta)}-\frac{a'(\eta)}{a(\eta)}\right]+\hbar^2\left[\frac{a''(\eta)}{2a(\eta)}-\frac{3 a'(\eta)^2}{4 a(\eta)^2}\right]\right)\psi_2(\eta)=&0
 ,\label{eq:DecoupledDirac2}
\end{align}
where we defined the the effective frequency
 \begin{align}
\omega_{k}(\eta)^2 := p(\eta)^2+m^2a(\eta)^2. \label{eq:omega}
\end{align}

%%%%%%%%%%%%%%%%%%%%%%%%%%%%%%%%%%%%%%%%%%%%%%%%%%%%%%%%%%%%%%%%%%%%%%%%%%%%%%%%%%%%%%%%%%%%%%%%%%%%%%%%%%%%%%%%%%%%%%%%%%%%%%%%%%%%
\section{Semiclassical saddlepoint method}
%%%%%%%%%%%%%%%%%%%%%%%%%%%%%%%%%%%%%%%%%%%%%%%%%%%%%%%%%%%%%%%%%%%%%%%%%%%%%%%%%%%%%%%%%%%%%%%%%%%%%%%%%%%%%%%%%%%%%%%%%%%%%%%%%%%%
\label{sec:SC}
In this section we compute the pair creation rate of general electric fields in \(\text{dS}_2\). This calculation is strongly inspired by the one proposed in \cite{us}, where the semiclassical number of pairs is computed in $\text{dS}_4$. It was remarked there that there is an interesting parallel between one-component fields in curved spacetime and two-component fields in flat spacetime, for which a generalization of the well known semiclassical techniques (see e.g.\cite{Brezin1970,Popov1971,Popov1972,Popov1973,Marinov1977,Popov2001,Kleinert2008,Kleinert2013,Strobel2014}) was found in \cite{Strobel2015}. This comes from the fact that both set-ups have two degrees of freedom. To make this parallel more obvious we now consider the field
\begin{align}
\Psi(x)=\sqrt{a(\eta)}\psi(x). \label{eq:spinor}
\end{align}
This can be seen as an equivalent of the  Mukhanov-Sasaki variable in inflation models \cite{RedBook}.
For this field the Dirac equation (\ref{eq:CoupledDirac1})-(\ref{eq:CoupledDirac2}) becomes an equation which is comparable to a Dirac equation in flat space-time with time dependent mass
\begin{align}
\label{to solve 1} & \ii\hbar \Psi'_1(\eta)+p(\eta) \Psi_1(\eta)-ma(\eta)\Psi_2(\eta)=0,\\ 
\label{to solve 2} & \ii\hbar \Psi'_2(\eta)-p(\eta) \Psi_2(\eta)-ma(\eta)\Psi_1(\eta)=0.
\end{align}
To obtain the number of created pairs we will now reformulate this equations as equations of the mode functions \(\alpha(\eta)\), \(\beta(\eta)\). To compute $|\beta(\eta\rightarrow0)|^2$ we will then perform a multiple integral iteration. Subsequently we will simplify these integrals using a a semiclassical saddle point approximation.\\
Inspired by similarities between Eqs.~(11)-(12) of \cite{Strobel2015} and (\ref{to solve 1})-(\ref{to solve 2}), we propose the following ansatz
\begin{align}\Psi_{1}(\eta)&=\frac{1}{\sqrt{2\omega_{k}(\eta)}}\left(\alpha(\eta)\sqrt{\omega_{k}(\eta)-p(\eta)}\,{\e^{-\frac{\ii}{2}K(\eta)}} +\beta(\eta)\sqrt{\omega_{k}(\eta)+p(\eta)}\,{\e^{\frac{\ii}{2}K(\eta)}}\right),\label{eq:ansatz1} \\
 \Psi_{2}(\eta)&=\frac{1}{\sqrt{2\omega_{k}(\eta)}}\left(\alpha(\eta)\sqrt{\omega_{k}(\eta)+p(\eta)}\,{\e^{-\frac{\ii}{2}K(\eta)}} -\beta(\eta)\sqrt{\omega_{k}(\eta)-p(\eta)}\,{\e^{\frac{\ii}{2}K(\eta)}}\right), \label{eq:ansatz2}
\end{align}
with the integral
 \begin{align}
 K(\eta)& \:= \frac{2}{\hbar} \int_{-\infty}^{\eta} \omega_{k}(\tau) d \tau
 \label{eq:K_xy}.
 \end{align}
 The Dirac equation (\ref{to solve 1})-(\ref{to solve 2}) leads to coupled differential equations for the mode functions by inserting the ansatz (\ref{eq:ansatz1})-(\ref{eq:ansatz2}). They read
\begin{align}
 \alpha'(\eta)&=\frac{\omega_{k}'(\eta)}{2 \omega_{k}(\eta)}G(\eta)\, \e^{\ii K(\eta)}\beta(\eta),\label{eq:alphaferm}\\
 \beta'(\eta)&=-\frac{\omega_{k}'(\eta)}{2 \omega_{k}(\eta)}G(\eta)\, \e^{-\ii K(\eta)}\alpha(\eta),\label{eq:betaferm}
\end{align}
with
\begin{align}
& G(\eta)=\frac{p(\eta)}{ma(\eta)}-\frac{\omega_{k}(\eta)p'(\eta)}{ma(\eta)\omega_{k}'(\eta)}, 
\end{align}
which can be seen as fermionic corrections to the analog bosonic case.\\
The initial conditions for the mode functions are chosen such that at past infinity there are only negative frequency modes, i.e.
\begin{align}
\alpha(-\infty)=1, && \beta(-\infty)=0. 
\end{align}
It is now possible to iteratively integrate  (\ref{eq:alphaferm}) and (\ref{eq:betaferm}), which leads to
\begin{equation}
 \begin{split}
 \beta(0)=&\sum_{m=0}^\infty (-1)^{m+1}\int_{-\infty}^0 d\eta_0\frac{\omega_k'(\eta_0)}{2\omega_k(\eta_0)}G(\eta_0)\e^{-iK(\eta_0)} \\&  \hspace{2cm}\times
\prod_{n=1}^m \int_{-\infty}^{\eta_{n-1}} d\tau_n\frac{\omega_k'(\tau_n)}{2\omega_k(\tau_n)}G(\tau_n)\e^{iK(\tau_n)} \int_{-\infty}^{\tau_n}d\eta_n\frac{\omega_k'(\eta_n)}{2\omega_k(\eta_n)} G(\eta_n)\e^{-iK(\eta_n)}. \\\label{eq:multiint}
 \end{split}
\end{equation}
One can use the fact these integrals are dominated by the classical turning points \cite{Berry1982}, which are given by 
\begin{align}
\label{eq:TP}
 \omega_k(\eta_p^\pm)=0 .
\end{align}
It is possible to show that for one pair of simple turning points the momentum spectrum of the pair creation rate in a semiclassical saddlepoint approximation is given by (see \cite{Strobel2015,Dumlu2011,Berry1982})
\begin{align}
 n_{k} = \lim_{\eta\rightarrow0} \left|\beta(\eta)\right|^2 = \left|\e^{-iK(\eta_p^{-})}\right|^2 \label{eq:MomentumSpectrumConst}.
\end{align}
The detailed intermediate steps of the derivation can be found in Eqs.~(32)-(38) of \cite{Strobel2015}. Observe that in \cite{Strobel2015,Dumlu2011,Berry1982} the integration contour is closed in the upper imaginary half plane whereas, we (as in \cite{us}) close it in the lower imaginary half plane because of opposite convention for the phases in (\ref{eq:multiint}).\\
The above result represents the semiclassical number of pairs created in each mode \(k\) for general electric fields in \(\text{dS}_2\). In the following sections we will concentrate on the constant electric field, where more explicit computations can be performed.

%%%%%%%%%%%%%%%%%%%%%%%%%%%%%%%%%%%%%%%%%%%%%%%%%%%%%%%%%%%%%%%%%%%%%%%%%%%%%%%%%%%%%%%%%%%%%%%%%%%%%%%%%%%%%%%%%%%%%%%%%%%%%%%%%%%%
\section{Constant electrical field in \texorpdfstring{\(\text{dS}_2\)}{dS2} spacetime}
%%%%%%%%%%%%%%%%%%%%%%%%%%%%%%%%%%%%%%%%%%%%%%%%%%%%%%%%%%%%%%%%%%%%%%%%%%%%%%%%%%%%%%%%%%%%%%%%%%%%%%%%%%%%%%%%%%%%%%%%%%%%%%%%%%%%
\label{sec:constantfield}
In this section we will study the Schwinger effect and the expectation value of the induced current in the vacuum at negative infinity. We first use the general method of the previous section to compute the semiclassical number of produced pairs per momentum mode \(k\) for this field configuration in section \ref{sec:constantSC}. We then construct positive and negative frequency mode solutions of the Dirac equation at past and future infinity in section \ref{sec:solutions}. These solutions are used to compute the number of produced pairs per momentum mode \(k\) using the Bogoliubov method in section \ref{sec:Bogoliubov}. We then compute the pair creation rate from the previous results in section \ref{sec:currentfromnk} before computing and regularizing the current in \ref{sec:current}.\\
The constant electric field in dS space is described by
\begin{align}
A(\eta)=\frac{E}{H^2 \eta}, \label{eq:constantfield}
\end{align}
since a comoving observer with a four velocity of \(u^{\mu}\) would measure the field
\begin{align}
E_{\mu} = u^{\nu} F_{\mu \nu}= a E\delta^{z}_{\mu},
\end{align}
which has constant field strength $E_{\mu}E^{\mu}=E^2$. \\
We introduce
\begin{align}
\lambda= \frac{eE}{H^2},&&\gamma= \frac{m}{H} \label{eq:parameters}
\end{align}
to characterize the electrical an gravitational field strength respectively. The effective frequency (\ref{eq:omega}) takes the form  
\begin{align}
 \omega_k(\eta)^2=k^2+2k\frac{\lambda}{\eta}+\frac{\lambda^2+\gamma^2}{\eta^2}. \label{eq:omegaexplicit}
\end{align}
Since we will later construct the positive and negative frequency solutions at past and future infinity we will here shortly comment on which form they should take. There form is given by the WKB solution
\begin{align}
 \psi^\pm(\eta)\sim \frac{1}{\sqrt{\omega_k(\eta)}}\exp\left(\pm\frac{\ii}{\hbar}\left|\int^\eta\omega_k(\tau)d\tau\right|\right).
\end{align}
Using the explicit form of the effective frequency given in (\ref{eq:omegaexplicit}) we find that the asymptotic behavior of the positive and negative frequency modes at past infinity, i.e.~for \(\eta\rightarrow- \infty\) is given by
\begin{align}
 \psi_\text{in}^\pm(\eta)\sim \frac{1}{\sqrt{|k|}}\exp\left(\mp\frac{\ii}{\hbar}|k|\eta\right). \label{eq:modein}
\end{align}
In the same way the behavior of positive and negative frequency modes at future infinity, i.e.~at \(\eta\rightarrow0\), is given by
\begin{align}
 \psi_\text{out}^\pm(\eta)\sim \frac{\eta^{\mp\mu+\frac{1}{2}}}{\sqrt{|\mu|}},\label{eq:modeout}
\end{align}
where we defined
\begin{align}
 \mu:=\frac{\ii}{\hbar}\sqrt{\gamma^2+\lambda^2}.
\end{align}

%%%%%%%%%%%%%%%%%%%%%%%%%%%%%%%%%%%%%%%%%%%%%%%%%%%%%%%%%%%%%%%%%%%%%%%%%%%%%%%%%%%%%%%%%%%%%%%%%%%%%%%%%%%%%%%%%%%%%%%%%%%%%%%%%%%%
\subsection{Semiclassical number of created pairs}
%%%%%%%%%%%%%%%%%%%%%%%%%%%%%%%%%%%%%%%%%%%%%%%%%%%%%%%%%%%%%%%%%%%%%%%%%%%%%%%%%%%%%%%%%%%%%%%%%%%%%%%%%%%%%%%%%%%%%%%%%%%%%%%%%%%%
\label{sec:constantSC}
Here we use the semiclassical saddlepoint method described in section \ref{sec:SC} to compute the number of pairs created by the constant electric field defined in (\ref{eq:constantfield}). We start by computing the turning points (\ref{eq:TP}), which are given by
\begin{align}
%\eta_p^{\pm}=\frac{-k\lambda\pm\ii\gamma\sqrt{k^2}}{k^2}.
\eta_p^{\pm}=-\frac{\lambda}{k}\pm\frac{\ii\gamma}{|k|},
\end{align}
where the absolute value of \(k\) was introduced such that \(\eta_p^-\) always denotes the turning point in the lower imaginary plane. We now find that the real part of \(\eta_p^{\pm}\) is only negative, i.e. inside the integration contour which is used for the approximation of (\ref{eq:multiint}), if \(k\) and \(\lambda\) have the same sign. In \cite{Froeb2014} this is called pair production in ``screening'' direction because the created pairs would reduce the electric field if we would allow backreaction. In the language of false vacuum decay this would be ``downward tunneling''. We find that as in flat spacetime in the semiclassical limit no pairs are produced in ``anti-screening'' direction which would be connected to ``upward'' tunneling. This however does no longer hold true when we compute the number of pairs using the Bogoliubov method in section \ref{sec:Bogoliubov}.\\
Only the imaginary part of $K(\eta_p^-)$ is contributing to (\ref{eq:MomentumSpectrumConst}). It is given by
\begin{align}
\Im[K(\eta_p^-)]=-\frac{\pi}{\hbar}\left(\sqrt{\gamma^2+\lambda^2}-|\lambda|\right)\theta(k \lambda),
\end{align}
where \(\theta(x)\) is the Heaviside step function.
Thus we find that number of pairs created per mode \(k\) (\ref{eq:MomentumSpectrumConst}) in the semiclassical limit is 
\begin{align}
n_{k}  = \exp\left[-\frac{2\pi}{\hbar} \left(\sqrt{\gamma^2+\lambda^2}-|\lambda|\right)\right]\theta(k \lambda). \label{eq:nkSC}
\end{align}
At this point it is insightful to discuss what semiclassical means in this context. The number of created pairs is related to the effective action via \(n_k= \exp(-S/\hbar)\). The semiclassical limit is the limit of large action, i.e.~$|\mu|-|\lambda| \gg 1$. A necessary condition for this to happen is $|\mu| \gg 1$. Two regimes can be discussed:
\paragraph*{Weak electric field:} For $\lambda \ll \gamma$, one finds $S \sim 2 \pi (\gamma - |\lambda|)$. The pairs are mainly produced by gravitational (cosmological) pair creation. The first term is the usual Boltzmann factor for non-relativistic massive particles at the Gibbons-Hawking temperature, while the second term is the correction of the small electric field. The number of produced particles gets suppressed by the electric field in this limit. 
\paragraph*{Strong electric field:}
For a strong electric field one finds $S= \pi m^2/(eE)$, which is the usual semiclassical action for the Schwinger effect in flat spacetime. The effect of curvature is negligible in this limit. \\
However for the constant field we are not limited to semiclassical calculations but can find the solutions of the Dirac equation explicitly.
 We will use this in the following to compute the number of pairs with the Bogoliubov method and compare it to the semiclassical result.
%%%%%%%%%%%%%%%%%%%%%%%%%%%%%%%%%%%%%%%%%%%%%%%%%%%%%%%%%%%%%%%%%%%%%%%%%%%%%%%%%%%%%%%%%%%%%%%%%%%%%%%%%%%%%%%%%%%%%%%%%%%%%%%%%%%%
\subsection{Solutions for the constant field}
%%%%%%%%%%%%%%%%%%%%%%%%%%%%%%%%%%%%%%%%%%%%%%%%%%%%%%%%%%%%%%%%%%%%%%%%%%%%%%%%%%%%%%%%%%%%%%%%%%%%%%%%%%%%%%%%%%%%%%%%%%%%%%%%%%%%
\label{sec:solutions}
Here we construct the positive and negative frequency solutions at past and future infinity for the constant field. Introducing the new variables 
\begin{align}
 z:=\frac{2\ii}{\hbar}k\eta, &&\kappa:=-\frac{\ii}{\hbar}\lambda, %&& r=\sgn(k),
\end{align}
in (\ref{eq:DecoupledDirac1}) and (\ref{eq:DecoupledDirac1}) for the constant field (\ref{eq:constantfield}) we find
\begin{align}
 \psi_1''(z)+\left(\frac{1}{z^2}\left[\frac{1}{4}-\mu^2\right]+\frac{1}{z}\left[\kappa-\frac{1}{2}\right]-\frac{1}{4}\right)\psi_1(z)=&0, \label{eq:Whittaker1}\\
 \psi_2''(z)+\left(\frac{1}{z^2}\left[\frac{1}{4}-\mu^2\right]+\frac{1}{z}\left[\kappa+\frac{1}{2}\right]-\frac{1}{4}\right)\psi_2(z)=&0.
 \label{eq:Whittaker2}
\end{align}
%\subsubsection{Solutions of the equations of motion}
Solutions of these equations are the Whittaker functions \(W_{\kappa\pm\frac12,\mu}(z),\,M_{\kappa\pm\frac12,\mu}(z)\) \cite{Olver2010}. We can make the following ansatz for two independent solutions of the decoupled Dirac equation (\ref{eq:DecoupledDirac1})-(\ref{eq:DecoupledDirac2})
\begin{align}
 \psi^a(z)=\begin{pmatrix}
            C_1\,W_{\kappa-\frac12,\mu}(z)\\
            C_2\,W_{\kappa+\frac12,\mu}(z)
           \end{pmatrix},&&
 \psi^b(z)=\begin{pmatrix}
            C_3\,W_{-\kappa+\frac12,-\mu}(-z)\\
            C_4\,W_{-\kappa-\frac12,-\mu}(-z) \label{eq:Csolutions}
           \end{pmatrix}.
\end{align}
Since we solved the two decoupled equations separately we lost the information about the coupling. It can be recovered by using the solutions (\ref{eq:Csolutions}) in one of the coupled equations (\ref{eq:CoupledDirac1})-(\ref{eq:CoupledDirac2}). Using the identity (\ref{eq:pmonehalfIdentity}) %in (\ref{eq:CoupledDirac1}) for \(\psi_1(z)\) and in (\ref{eq:CoupledDirac2}) for \(\psi_2(z)\) respectively, 
we find
\begin{align}
  C_{\nicefrac{1}{4}}=-\sqrt{\mu^2-\kappa^2}\,C_{\nicefrac{2}{3}}=-\frac{\ii}{\hbar}\gamma \,C_{\nicefrac{2}{3}}. \label{eq:C_23}
\end{align}
We can show with the help of (\ref{eq:Wconjugated}) that with this values for the constants \(C_{\nicefrac{2}{3}}\)  the solutions (\ref{eq:Csolutions}) are orthogonal
\begin{align}
 \psi^a(z)^\dagger\cdot\psi^b(z)=0. \label{eq:orthogonality}
\end{align}
The value of the remaining constants will be found by asking for normalization after quantizing.\\
%\subsubsection{Construction of positive and negative frequency solutions}
We can now use the limit of the function \(W_{\kappa,\mu}(z)\) for \(|z|\rightarrow\infty\) given in (\ref{eq:Wlimit}) to find the behavior of the solutions (\ref{eq:Csolutions}) at past infinity
\begin{align}
 \lim_{\eta\rightarrow-\infty}\psi^a(z)=\lim_{\eta\rightarrow-\infty}
 \begin{pmatrix}
  C_1\,z^{\kappa-\frac{1}{2}}\\
  C_2\,z^{\kappa+\frac{1}{2}}
 \end{pmatrix} \exp\left(-\frac{\ii}{\hbar}k\eta\right),\\
 \lim_{\eta\rightarrow-\infty}\psi^b(z)=\lim_{\eta\rightarrow-\infty}
 \begin{pmatrix}
  C_3\,(-z)^{-\kappa+\frac{1}{2}}\\
  C_4\,(-z)^{-\kappa-\frac{1}{2}}
 \end{pmatrix} \exp\left(\frac{\ii}{\hbar}k\eta\right).
\end{align}
Comparing this to the desired asymptotic behavior of the mode functions given in (\ref{eq:modein}) we find that the positive frequency solution \(\psi_\text{in}^+(\eta)\) is given by \(\psi^a(z)\) for \(k>0\) and by \(\psi^b(z)\) for \(k<0\). Thus the positive and negative frequency solutions in the asymptotic past  can be constructed as
\begin{align}
 \psi_\text{in}^+(z)=\begin{cases}\psi^a(z)& k>0\\
                        \psi^b(z)& k<0
              \end{cases},&&
\psi_\text{in}^-(z)=\begin{cases}\psi^b(z)& k>0\\
                        \psi^a(z)& k<0
              \end{cases}.&& \label{eq:2DFermSol}
\end{align}
Following from (\ref{eq:orthogonality}) also these solutions are orthogonal.\\
We can now construct the spinor field operator by specifying the momentum decomposition (\ref{eq:decomp1})   
\begin{align}
 \psi(x)=\int \frac{dk}{2\pi\hbar}\e^{\frac{\ii}{\hbar} k x_1}\left[b(k) \psi^+(\eta)+d^\dagger(-k)\psi^-(\eta)\right].\label{eq:decomp2}
\end{align}
We impose canonical anti-commutation relations
\begin{align}
 \left\{\psi_\alpha(x_1,\eta),\pi(x_1',\eta)\right\}=\ii\hbar\,\delta(x_1-x_1')\delta_{\alpha\beta}. 
\end{align}
Using the conjugate momentum (\ref{eq:mometum}) in dS spacetime we find that this is equivalent to
\begin{align}
\left\{\psi_\alpha(x_1,\eta),\psi_\beta^\dagger(x_1',\eta)\right\}=\frac{1}{a(\eta)}\delta(x_1-x_1')\delta_{\alpha\beta}. \label{eq:comm}
\end{align}
This holds true if the creation and annihilation operators follow the anti-commutation relations
\begin{align}
 \left\{b(k),b(k')^\dagger\right\}=\left\{d(k),d(k')^\dagger\right\}= 2\pi \hbar\,\delta(k-k'). \label{eq:comm2}
\end{align}
 and the mode functions fulfill the Wronskian condition
\begin{align}
\psi^+(\eta)\psi^+(\eta)^\dagger+\psi^-(\eta)\psi^-(\eta)^\dagger=\frac{1}{a(\eta)}\mathbbm{1}.\label{eq:WronskianCondition}
\end{align}
This is true for the solutions (\ref{eq:2DFermSol}) if (see App.~\ref{sec:WronskianCondition})
\begin{align}
 C_2=C_3%=\sqrt{\frac{\hbar}{2va(\eta)}}\e^{\frac{\pi}{2}\ii\kappa\sgn(k)}
 =\sqrt{\frac{\hbar H}{2|k|}}\e^{\frac{\pi}{2}\ii\kappa\sgn(k)} \label{eq:C_14}.
\end{align}
Accordingly we find that the positive and negative frequency solutions at asymptotic past infinity are given by
\begin{align}
 \psi_\text{in}^+(\eta)=\sqrt{\frac{\hbar H}{2|k|}}\begin{cases}
               \e^{\frac{\pi}{2}\ii\kappa}
               \begin{pmatrix}
                \frac{\gamma}{\ii\hbar}\,W_{\kappa-\frac12,\mu}(2\ii v)\\
                 \,W_{\kappa+\frac12,\mu}(2\ii v)
               \end{pmatrix}& %k>0
               \\
               {\e^{-\frac{\pi}{2}\ii\kappa}}\begin{pmatrix}
                \,W_{-\kappa+\frac12,-\mu}(2\ii v)\\
                \frac{\gamma}{\ii\hbar}\,W_{-\kappa-\frac12,-\mu}(2\ii v)
               \end{pmatrix}&%k<0
              \end{cases},&&
\psi_\text{in}^-(\eta)=\sqrt{\frac{\hbar H}{2|k|}}\begin{cases}
               {\e^{\frac{\pi}{2}\ii\kappa}}
               \begin{pmatrix}
                \,W_{-\kappa+\frac12,-\mu}(-2\ii v)\\                
               \frac{\gamma}{\ii\hbar}\,W_{-\kappa-\frac12,-\mu}(-2\ii v)
               \end{pmatrix}& k>0\\
               {\e^{-\frac{\pi}{2}\ii\kappa}}\begin{pmatrix}
                \frac{\gamma}{\ii\hbar}\,W_{\kappa-\frac12,\mu}(-2\ii v)\\
                \,W_{\kappa+\frac12,\mu}(-2\ii v)
               \end{pmatrix}&k<0
              \end{cases}, \label{eq:2DFermSol2}
\end{align}
where we introduced \(v:=|k|\eta/\hbar\).
In an analogous way we can construct the positive and negative frequency solutions at \(\eta\rightarrow0\) as
\begin{align}
 \psi^+_\text{out}(\eta)=\frac{1}{2}\sqrt{\frac{\hbar H}{|k|}}%\begin{cases}
               \e^{\frac{\pi}{2}\ii r\mu }
               \begin{pmatrix}
                 \sqrt{\frac{\mu-\kappa}{\mu}}\,M_{\kappa-\frac12,\mu}(z)\\
                 \sqrt{\frac{\mu+\kappa}{\mu}}\,M_{\kappa+\frac12,\mu}(z)
               \end{pmatrix} 
,&&
 \psi_\text{out}^-(\eta)=\frac{1}{2}\sqrt{\frac{\hbar H}{|k|}}           
              %\begin{cases}
               \e^{\frac{\pi}{2}\ii r\mu }
                \begin{pmatrix}
                 \sqrt{\frac{\mu+\kappa}{\mu}}\,M_{-\kappa+\frac12,-\mu}(-z)\\
                 -\sqrt{\frac{\mu-\kappa}{\mu}}\,M_{-\kappa-\frac12,-\mu}(-z)
                \end{pmatrix},
\label{eq:2DFermSolout}
\end{align}
where we introduced 
\begin{align}
 r:=\sgn(k).
\end{align}
Using the limit of the function \(M_{\kappa,\mu}(z)\) for \(z\rightarrow0\) given by (\ref{eq:Mlimit}) we find
\begin{align}
 \lim_{\eta\rightarrow0}\psi^+_\text{out}(\eta)&=\lim_{\eta\rightarrow0}\frac{1}{2}\sqrt{\frac{\hbar H}{|k|}}
               \e^{\ii\frac{\pi}{2}r\mu}
               \begin{pmatrix}
                 \sqrt{\frac{\mu-\kappa}{\mu}}\\
                 \sqrt{\frac{\mu+\kappa}{\mu}}
               \end{pmatrix}\left(\ii\frac{2}{\hbar}|k|\eta\right)^{\mu+\frac{1}{2}}
,\\
 \lim_{\eta\rightarrow0} \psi_\text{out}^-(\eta)&=
 \lim_{\eta\rightarrow0}\frac{1}{2}\sqrt{\frac{\hbar H}{|k|}}\e^{\ii\frac{\pi}{2}r\mu}
                \begin{pmatrix}
                 \sqrt{\frac{\mu+\kappa}{\mu}}\\
                 -\sqrt{\frac{\mu-\kappa}{\mu}}
                \end{pmatrix}\left(-\ii\frac{2}{\hbar}|k|\eta\right)^{-\mu+\frac{1}{2}}.
\end{align}
Comparing to (\ref{eq:modeout}) one finds that the modes %(\ref{eq:2DFermSolout1})-
(\ref{eq:2DFermSolout}) have the right asymptotic behavior. They can also be shown to follow the Wronskian condition (\ref{eq:WronskianCondition}) by performing steps analogous to the ones found in appendix \ref{sec:WronskianCondition}.\\
Observe that the positive and negative frequency also correspond to the particle and antiparticle solution. This can be seen by defining the charge conjugate spinor representing the antiparticle of $\psi(x)$ as $\psi^c(x) := \ii \sigma_2 \psi^*(x)$, with the Pauli matrix 
\begin{align}
 \sigma_2=\begin{pmatrix}
           0&-\ii\\
           \ii&0
          \end{pmatrix}.
\end{align}
This is a physical illustration of Feynman's picture that antiparticles are traveling backwards in time \cite{PhysRev.76.749}.
%%%%%%%%%%%%%%%%%%%%%%%%%%%%%%%%%%%%%%%%%%%%%%%%%%%%%%%%%%%%%%%%%%%%%%%%%%%%%%%%%%%%%%%%%%%%%%%%%%%%%%%%%%%%%%%%%%%%%%%%%%%%%%%%%%%%
\subsection{Number of pairs using the Bogoliubov method}
%%%%%%%%%%%%%%%%%%%%%%%%%%%%%%%%%%%%%%%%%%%%%%%%%%%%%%%%%%%%%%%%%%%%%%%%%%%%%%%%%%%%%%%%%%%%%%%%%%%%%%%%%%%%%%%%%%%%%%%%%%%%%%%%%%%%
\label{sec:Bogoliubov}
We now will use the method of Bogoliubov coefficients to compute the number of created pairs per mode \(k\) in dS spacetime. 
Similar methods have been used to compute the pair creation rate in time-dependent fields in flat spacetime for general D-dimensional fields in \cite{Gavrilov1996}, without an electric field for bosons in dS in \cite{Anderson2014} and for the constant field in \(\text{dS}_2\) spacetime for fermions and bosons respectively in \cite{Haouat2013} and \cite{Froeb2014}. In \cite{Kluger1998} the connection of this technique to kinetic theory was shown in the bosonic case.
\\
We are considering cases where there is a vacuum state for the produced particles in the asymptotic future. This requires the rate of change of the background to be small in the asymptotic future, that is
\begin{equation}
\left(\frac{\omega'_{k}(\eta)}{\omega_{k}^2(\eta)}\right)^2 \underset{\eta \rightarrow 0}{\sim} |\mu|^{-2} \text{ and } \left(\frac{\omega''_{k}(\eta)}{\omega_{k}^3(\eta)}\right) \underset{\eta \rightarrow 0}{\sim} 2|\mu|^{-2}\end{equation}
being small. We see that this is the case when $|\mu| \gg 1$.\\
To use the the method of Bogoliubov coefficients we use the fact that the positive frequency mode at past infinity is connected to the modes at \(\eta\rightarrow0\) through
\begin{align}
 \psi_\text{in}^+(\eta)=\alpha_k \,\psi_\text{out}^+(\eta)+\beta_k\, \psi_\text{out}^-(\eta) \label{eq:inout},
\end{align}
where the Bogoliubov coefficients are normalized as
\begin{align} 
 |\alpha_k|^2+|\beta_k|^2=1.
\end{align}
The coefficients can now be found by putting the explicit form of the solutions (\ref{eq:2DFermSol2}) and %(\ref{eq:2DFermSolout1})-
(\ref{eq:2DFermSolout}) in  (\ref{eq:inout}) and using the connection between the Whittaker functions \(W_{\kappa,\mu}(z)\) and \(M_{\kappa,\mu}(z)\) given in (\ref{eq:WtoM}). Using (\ref{eq:Mproperty}) this leads to 
\begin{align}
\alpha_k=\frac{\Gamma(-2\mu)}{\Gamma(-\mu- r \kappa)}\frac{\sqrt{2\mu}}{\sqrt{\mu+ r \kappa}}\e^{-\frac{\pi}{2}\ii(\mu- r \kappa)}\e^{\frac{\pi}{4}\ii(r-1)}
,&&
\beta_k=\frac{\Gamma(2\mu)}{\Gamma(\mu- r \kappa)}\frac{\sqrt{2\mu}}{\sqrt{\mu- r \kappa}}\e^{\frac{\pi}{2}\ii(\mu+ r \kappa)}\e^{\frac{\pi}{4}\ii(r+1)}.
\end{align}
We thus find the number of created pairs per mode to be
\begin{align}
 n_k=|\beta_k|^2=\e^{-\pi(|\mu|-\ii r \kappa)}\frac{\sinh(\pi(|\mu|+\ii r \kappa))}{\sinh(2\pi |\mu|)}\label{eq:nk}.
\end{align}
This result can be shown to be equivalent to Eq.~(19) of  \cite{Haouat2013}, where it has been derived in an equivalent way. Comparing it to the bosonic result in \(\text{dS}_2\) of \cite{Froeb2014} (see Eq.~(2.18)) we find that the only difference is a \(\sinh(\pi(|\mu|-  \ii r \kappa))\) instead of a \(\cosh(\pi(|\mu|-\ii r \kappa))\).  However this difference vanishes in the relevant limit $|\mu| \gg 1$.\\
Comparing to the semiclassical result given in (\ref{eq:nkSC}) the most striking difference is that according to (\ref{eq:nk}) particles can also be produced in ``anti-screening'' direction which corresponds to ``upward'' tunneling. This was already found in the bosonic case \cite{Froeb2014}. However in the limit, i.e.~\(|\mu|\gg1, \lambda\gg\hbar\), the two results agree, since pair production in ``anti-screening'' direction gets exponentially suppressed.
%%%%%%%%%%%%%%%%%%%%%%%%%%%%%%%%%%%%%%%%%%%%%%%%%%%%%%%%%%%%%%%%%%%%%%%%%%%%%%%%%%%%%%%%%%%%%%%%%%%%%%%%%%%%%%%%%%%%%%%%%%%%%%%%%%%%
\subsection{Pair creation rate from the number of pairs}
%%%%%%%%%%%%%%%%%%%%%%%%%%%%%%%%%%%%%%%%%%%%%%%%%%%%%%%%%%%%%%%%%%%%%%%%%%%%%%%%%%%%%%%%%%%%%%%%%%%%%%%%%%%%%%%%%%%%%%%%%%%%%%%%%%%%
\label{sec:currentfromnk}
One can compute the pair creation rate from the number of pairs per momentum mode \(k\) by integrating
\begin{equation}
\Gamma:= \frac{1}{ V} \int \frac{d k}{2\pi \hbar}\, n_{k}
\end{equation} \\
where $ V:= a(\eta)^2 d\eta$ is the unit two volume of the spacetime. Here we use a derivation of the number of pairs created, which  relies on an estimate for the moment when most of the particles are created. 
 A good estimate for the moment when the particles are created is given when the violation of the adiabaticity is maximal, i.e.~when the rate of change of the frequency $\omega_{k}$ is extremal. 
 One can show that the maximum gives the estimate for the creation time \cite{Froeb2014} 
\begin{equation}
\label{estimate}
\eta \sim -\frac{|\mu|}{|k|}\hbar.
\end{equation}
Using this the \(k\)-integral can be changed into a time integral. Detailed justification of this procedure can be found in\cite{Gavrilov1996,Kluger1998,Anderson2014}. The pair production rate can thus be estimated from (\ref{eq:nk}) by 
\begin{equation}
\Gamma \approx \frac{ |\mu|  H^2}{2\pi}  \frac{\cosh\left(\frac{2 \pi}{\hbar} \lambda\right)-\e^{-2\pi|\mu|}}{\sinh(2 \pi |\mu|)}. \label{eq:paircreationrate}
\end{equation}
We can also compute the physical number density \(n\) of produced pairs at the  time $\eta$ with the help of
\begin{equation}
n=\frac{1}{a(\eta)} \int_{-\infty}^{\eta}\, d\tau a(\tau)^2 \Gamma= \frac{\Gamma}{H}.
\end{equation}
As was already found for bosons this is constant, which shows that the particles created via Schwinger and gravitational particle creation compensate the effects of the expansion of the spacetime. That means that the fermion population is always dominated by the particles created within one Hubble time, in the limit where our approximations hold.\\
Performing calculations in analogy with the ones performed in  \cite{Kobayashi2014} one finds that in the limit of flat space-time, i.e. $H\rightarrow0$, one recovers the familiar results for Schwinger pair production in Minkowski spacetime (see e.g.~\cite{Gavrilov1996})
\begin{equation}
\label{M4}
 \lim_{H\rightarrow 0} \Gamma = \frac{1}{\hbar}\frac{|eE|}{ \pi} \exp \left(-\frac{\pi}{\hbar}\frac{ m^2}{|eE|} \right). \\
\end{equation}
We will finish this section by computing the vacuum decay rate. This rate was computed from the imaginary part of the one-loop effective action by Schwinger \cite{Schwinger1951}. It has been shown that Schwinger's result agrees with the canonical method for the case of pure electric field \cite{Nikishov1969,Naro1970,Fradkin1991} as well as in de Sitter space \cite{Mottola:1984ar}. In our case, the vacuum decay rate is defined as
\begin{equation}
\Upsilon=\frac{1}{ V} \int \frac{dk}{2 \pi \hbar} \log(1-n_k).
\end{equation}
Expanding the logarithm and changing the \(k\)-integral to a time integral, as above, it is possible to find the following expression for the decay rate
\begin{equation}
\Upsilon= \sum_{r=\pm 1}r \sum_{j=1}^{\infty}\frac{ |\mu|  H^2}{2\pi j}  \e^{-\pi j( |\mu| \ - \frac{\lambda r}{\hbar}) } \frac{\sinh^j(\pi (|\mu|+\frac{\lambda r}{\hbar})}{\sinh^j(2\pi |\mu|)}.
\end{equation}
Taking $H\rightarrow0$ gives the correct flat spacetime expression
\begin{equation}
 \lim_{H\rightarrow 0} \Upsilon =\frac{1}{\hbar} \sum_{j=1}^{\infty}\frac{|eE|}{ \pi j}  \exp \left(-j\frac{\pi}{\hbar} \frac{m^2}{|eE|} \right).
\end{equation}
%%%%%%%%%%%%%%%%%%%%%%%%%%%%%%%%%%%%%%%%%%%%%%%%%%%%%%%%%%%%%%%%%%%%%%%%%%%%%%%%%%%%%%%%%%%%%%%%%%%%%%%%%%%%%%%%%%%%%%%%%%%%%%%%%%%%
\subsection{Computation of the current}
%%%%%%%%%%%%%%%%%%%%%%%%%%%%%%%%%%%%%%%%%%%%%%%%%%%%%%%%%%%%%%%%%%%%%%%%%%%%%%%%%%%%%%%%%%%%%%%%%%%%%%%%%%%%%%%%%%%%%%%%%%%%%%%%%%%%
\label{sec:current}
In this section we will compute the expectation value of the current in a locally inertial coordinate system given by \(\underline{\gamma}^\mu(x)=\gamma^\mu\). {This coordinates  were introduced to make the probability density positive semi-definite at each spacetime point \cite{Pollock2010,parker1980one,parker1980one2}.  The expectation value of the current with respect to vacuum at past infinity is then given by 
\begin{align}
 J^x&=-\frac{e}{2}\left\langle0\right|\left[\overline{\psi}(x),{\gamma}^x\psi(x)\right] \left|0\right\rangle\\
 &=-\frac{e}{2}\int_{-\infty}^{\infty}\frac{dk}{2\pi\hbar}\left[-\psi_\text{in}^+(\eta)^\dagger\gamma^0\gamma^x\psi_\text{in}^+(\eta)+\psi_\text{in}^-(\eta)^\dagger\gamma^0\gamma^x\psi_\text{in}^-(\eta)\right],\\
 &=-\frac{e}{2}\int_{-\infty}^{\infty}\frac{dk}{2\pi\hbar}\left[|\psi_1^+(\eta)|^2-|\psi_2^+(\eta)|^2-|\psi_1^-(\eta)|^2+|\psi_2^-(\eta)|^2\right]. \label{eq:current1}
\end{align}
where \(\psi_1^\pm(\eta)\) and \(\psi_2^\pm(\eta)\) are the first and second component of \(\psi_\text{in}^\pm\) respectively. 
Using the diagonal elements of the Wronskian condition (\ref{eq:WronskianCondition}) we find a connection between the absolute square of the positive and negative frequency modes which can be used to simplify the current to
\begin{align}
 J^x=-e\int_{-\infty}^{\infty}\frac{dk}{2\pi\hbar}\left(|\psi_1^+(\eta)|^2-|\psi_2^+(\eta)|^2\right). \label{eq:current2}
\end{align}
Using the explicit form of the positive and negative frequency modes (\ref{eq:2DFermSol2}) this can be computed as (see Appendix \ref{sec:Integral}) 
\begin{align}
 J^x=&\frac{eH}{\pi}\ii\left(\mu\frac{\sin(2\pi\kappa)}{\sin(2\pi\mu)}-\kappa\right)\label{eq:explicitcurrent}.
\end{align}
We will now regularize this current using adiabatic subtraction. %%%%%%%%%%%%%%%%%%%%%%%%%%%%%%%%%%%%%%%%%%%%%%%%%%%%%%%%%%%%%%%%%%%%%%%%%%%%%%%%%%%%%%%%%%%%%%%%%%%%%%%%%%%%%%%%%%%%%%%%%%%%%%%%%%%%
\subsubsection{Adiabatic Regularization}
%%%%%%%%%%%%%%%%%%%%%%%%%%%%%%%%%%%%%%%%%%%%%%%%%%%%%%%%%%%%%%%%%%%%%%%%%%%%%%%%%%%%%%%%%%%%%%%%%%%%%%%%%%%%%%%%%%%%%%%%%%%%%%%%%%%%
Various methods exist to regularize and renormalize physical quantities. To name some of them, there are proper-time regularization, dimensional regularization, zeta-function regularization, Pauli-Villars subtraction, point splitting regularization (in particular by the Hadamar method) and adiabatic regularization (or subtraction). We propose to implement the last  one for our problem as it has been done in order regularize the current in the bosonic case in \(\text{dS}_4\) \cite{Kobayashi2014} as well as for fermions in flat Minkowski space in \cite{Kluger1992}.\\
Adiabatic regularization was first introduced by Parker to cure the UV divergence and the rapid oscillation of the particle number operator \cite{Parker1966}. Parker and Fulling generalized it to take care of the UV divergences of the stress energy tensor of scalar fields in homogeneous cosmological backgrounds \cite{Parker1974,Fulling1974,Fulling1974B}.\\
To regularize the current using adiabatic subtraction we compute the current for slow background variations and subtract it from our result (\ref{eq:explicitcurrent}). To quantify what is meant by slow varying background more precisely, we introduce a dimensionless slowness parameter T by replacing the scale factor \(a(\eta)\) by a family of functions $a_T (\eta) := a(\eta/T )$. Observe that in the limit of infinitely slow backgrounds, $T \rightarrow \infty$, the derivatives of \(a(\eta)\) will tend to zero since for integer $n$ 
\begin{align}\frac{d^n a(\eta/T)}{d\eta^n} \propto \frac{1}{T^n}.\end{align}
We define adiabatic orders as powers of $T^{-1}$. In our problem, it is equivalent to count time derivatives and adiabatic orders in a given expression. One of the reasons that we did not set $\hbar$ to one here is that since conformal time derivative are always multiplied by $\hbar$, it is actually possible to ``formally'' pose $T := 1/\hbar$ and to work in powers of $\hbar$.\\
For our purpose, we will expand our modes up to second adiabatic order and subtract it from the current to regularize it. To do so we start from the WKB-like ansatz
\begin{align}
 \psi^+_1(\eta)=N_1\sqrt{\frac{1}{2\Omega(\eta)}}\exp\left(\int^\eta\left[-\frac{\ii}{\hbar}\Omega(t)+\frac{p(t)}{2\Omega(t)}\left(\frac{p'(t)}{p(t)}-\frac{a'(t)}{a(t)}\right)\right]dt\right). \label{eq:adiabaticansatz}
\end{align}
Observe that in difference to the bosonic case of \cite{Kobayashi2014} we are not using a pure WKB ansatz. It was found that for fermions, a new kind of ansatz has to be proposed so that the imaginary part of the decoupled Dirac equation (\ref{eq:DecoupledDirac1}) is canceled. This has been used for fermions under the influence of an electric field in flat spacetime in \cite{Kluger1992} and in curved (Friedman-Lema{\^\i}tre-Robertson-Walker) spacetimes without electric field \cite{Landete2014} (see also \cite{Ghosh:2015mva}). The ansatz (\ref{eq:adiabaticansatz}) is a combination of the two previous ones. By putting the ansatz in the decoupled Dirac equation (\ref{eq:DecoupledDirac1}) we find a reparametrization of it in terms of \(\Omega(\eta)\), namely
\begin{align}
\begin{split}
 \Omega(\eta)^2-\omega(\eta)^2=&\hbar^2\left[\frac{a''(\eta)}{2 a(\eta)}\left(1+\frac{p(\eta)}{\Omega(\eta)}\right)-\frac{a'(\eta)^2}{a(\eta)^2}\left(\frac{3}{4}+\frac{p(\eta)}{2\Omega(\eta)}-\frac{p(\eta)^2}{4\Omega(\eta)^2}\right)+\frac{a'(\eta)}{2 a(\eta)}\frac{p'(\eta)}{\Omega(\eta)}\left(1-\frac{p(\eta)}{\Omega(\eta)}\right)\right.\\
 &\hspace{2cm}\left.+\frac{\Omega'(t)p(t)}{\Omega(t)^2}\left(\frac{p'(t)}{p(t)}-\frac{a'(t)}{a(t)}\right)-\frac{p''(\eta)+\Omega''(\eta)}{2\Omega(\eta)}+\frac{3\Omega'(\eta)^2+p'(\eta)^2}{4\Omega(\eta)^2}\right]. \label{eq:Omega-omega}
\end{split}
 \end{align}
We can now expand (\ref{eq:Omega-omega}) to find \(\Omega(t)=\omega(t)+\O(T)^{-2}\). The \((n+1)\)-th order can be found by iteratively using the \(n\)-th order solution on the right hand side of (\ref{eq:Omega-omega}).\\ 
For the second component of the spinor we use the ansatz (\ref{eq:adiabaticansatz}) in the coupled Dirac equation (\ref{eq:CoupledDirac1}) to find
\begin{align}
 \frac{\psi^+_2(\eta)}{\psi_1^+(\eta)}=\frac{\Omega(\eta)+p(\eta)}{m a(\eta)}-\frac{\ii \hbar}{2m a(\eta)} \left[\frac{\Omega'(\eta)+p'(\eta)}{\Omega(\eta)}-\frac{a'(\eta)}{a(\eta)}\left(1+\frac{p(\eta)}{\Omega(\eta)}\right)\right]\label{eq:adiabaticansatz2}.
\end{align}
Solving the off-diagonal element of the Wronskian condition (\ref{eq:WronskianCondition}) for \(\psi_1^-(\eta)\) and using it in one of the diagonal elements we can show that  
\begin{align}
 |\psi_1^+(\eta)|^2+|\psi_2^+(\eta)|^2=\frac{1}{a(\eta)}. \label{eq:normalization}
\end{align}
Using this normalization condition we can now write the current (\ref{eq:current2}) in terms of the fraction (\ref{eq:adiabaticansatz2})
\begin{align}
 J^x=-\frac{e}{a(\eta)}\int_{-\infty}^{\infty}\frac{dk}{2\pi\hbar}\frac{1- \frac{|\psi^+_2(\eta)|^2}{|\psi^+_1(\eta)|^2}}{ 1+\frac{|\psi^+_2(\eta)|^2}{|\psi^+_1(\eta)|^2}}. \label{eq:currentinbetween}
\end{align}
To perform the adiabatic expansion we can use the fact that \(\Omega(\eta)^2-\omega(\eta)^2\) is of second adiabatic order, which follows from (\ref{eq:Omega-omega}). It is possible to write \(\Omega(\eta)=\sqrt{\omega(\eta)^2+[\Omega(\eta)^2-\omega(\eta)^2]}\) and then expand the square root to find
\begin{align}
 \Omega(\eta)=\omega(\eta)+\frac{\Omega(\eta)^2-\omega(\eta)^2}{2\,\omega(\eta)}+\O(T)^{-4}.
\end{align}
Using this we can expand (\ref{eq:currentinbetween}) to second adiabatic order
\begin{align}
  J^x=\frac{e}{a(\eta)}\int_{-\infty}^{\infty}\frac{dk}{2\pi\hbar}\left(\frac{p(\eta)}{\omega(\eta)}+\frac{\omega(\eta)-p(\eta)}{2\omega(\eta)^3}\left[\Omega(\eta)^2-\omega(\eta)^2\right]+\frac{\hbar^2}{8}\frac{\left[ma'(\eta)p(\eta)-ma(\eta)p'(\eta)\right]^2}{\omega(\eta)^6}+\O\left(T\right)^{-4}\right),
\end{align}
where for \(\Omega(\eta)^2-\omega(\eta)^2\) one can use (\ref{eq:Omega-omega}) and replace \(\Omega(\eta)\) by \(\omega(\eta)\) on the right hand side. We can now compute this for the constant electric field and find %\(actually I have to introduce a \(\sgn(k)\) for this to be true)
\begin{align}
  J^x=\frac{e}{a(\eta)}\int_{-\infty}^{\infty}\frac{dk}{2\pi\hbar}\frac{p(\eta)}{\omega(\eta)}+\O\left(T\right)^{-4}=-\frac{eH}{\pi}\frac{\lambda}{\hbar}+\O\left(T\right)^{-4}.
\end{align}
For the bosonic case, the same counter term was found using the Pauli-Villars regularization method in $\text{dS}_2$ in \cite{Froeb2014}. The adiabatic regularization for bosons, in $\text{dS}_2$, is very similar to the calculation presented above and the final result can be found to be the same.\\
Performing the adiabatic subtraction for the current (\ref{eq:explicitcurrent}) we find 
\begin{align}
 J^x_\text{reg}=&\frac{eH}{\pi}\mu\frac{\sinh\left(\frac{2\pi}{\hbar}\lambda\right)}{\sin(2\pi\mu)}.\label{eq:regcurrent}
\end{align}
In the next subsections, we will discuss the properties of this current in detail. Performing the limit of strong electrical and strong gravitational field we underline the effect of the respective contributions to the total Schwinger effect. For a plot of the current (\ref{eq:regcurrent}) as a function of \(\lambda\) for different values of \(\gamma\) see Fig.~\ref{fig:current}. There we also show a comparison to the bosonic current of \cite{Froeb2014} which is discussed in section \ref{sec:comparison} and plotted as dotted lines.
%%%%%%%%%%%%%%%%%%%%%%%%%%%%%%%%%%%%%%%%%%%%%%%%%%%%%%%%%%%%%%%%%%%%%%%%%%%%%%%%%%%%%%%%%%%%%%%%%%%%%%%%%%%%%%%%%%%%%%%%%%%%%%%%%%%%
 \subsubsection{Strong and weak electric field limit}
%%%%%%%%%%%%%%%%%%%%%%%%%%%%%%%%%%%%%%%%%%%%%%%%%%%%%%%%%%%%%%%%%%%%%%%%%%%%%%%%%%%%%%%%%%%%%%%%%%%%%%%%%%%%%%%%%%%%%%%%%%%%%%%%%%%%
If we look at the limit \(\lambda\rightarrow \infty\) for fixed \(\gamma\), we find that the current is dominated by the electric field and has the asymptotic behavior
\begin{align}
 J^x_\text{reg}\sim \frac{eH}{\pi} \frac{\lambda}{\hbar}, \label{eq:biglambda}
\end{align}
which is independent of the mass. This linear behavior can also be seen in Fig.~\ref{fig:current} where all the curves align for large \(\lambda\). In the bosonic case in \(\text{dS}_2\) this linear behavior is also present \cite{Froeb2014} whereas in \(\text{dS}_4\) a quadratic behavior is found which leads to a linear behavior for the conductivity (defined as ${J}/{\lambda}$). This can be used to impose strong constrains on magnetogenesis scenarios \cite{Kobayashi2014}. For fermions in \(\text{dS}_4\) the same might appear. The strong field limit is the same limit as the small mass limit \(\gamma\ll\hbar\) for which we find
\begin{align}
 J^x_\text{reg}=\frac{eH}{\pi}\left(\frac{\lambda}{\hbar}+\left[\frac{\hbar}{2\lambda}-\pi\coth\left(\frac{2\pi}{\hbar}\lambda\right)\right]\frac{\gamma^2}{\hbar^2}\right)+\O\left(\frac{\gamma}{\hbar}\right)^3 \label{eq:smallmasslimit}.
\end{align}
For \(\gamma=0\) we find exactly the linear response which is analog to the asymptotic behavior (\ref{eq:biglambda}) since for strong enough fields the effect of the mass is negligible. The linear response is the usual flat spacetime response.} Indeed the Schwinger pair creation rate for massless carriers in flat spacetime is given by
\begin{equation}
\Gamma=\frac{eE}{2\pi\hbar},
\end{equation} which leads to the induced current
\begin{equation}
J=\frac{e^2 Et}{\pi\hbar}.
\end{equation} In an expanding spacetime the current can be computed in a comoving frame but will then be diluted in the physical frame. Naively making the substitution $t \rightarrow 1/H$, gives the linear response found in (\ref{eq:biglambda}). This substitution will be used again in the context of the flat spacetime limit in section \ref{sec:wog}. This shows that taking the limit $\gamma = m/H \rightarrow 0$ is equivalent to take the limit $H \rightarrow 0$ and $m=0$ which is the flat spacetime limit for massless particles. This illustrates the fact that a massless fermion is conformally invariant.\\
The limit of strong electric fields also allows us to compare to the result of the pair creation rate found in section \ref{sec:currentfromnk}. An approximation of the current for relativistic particles is given by
\begin{align}
 J\approx2en=2e\frac{\Gamma}{H}. \label{eq:approx_current}
\end{align}
Since the particle picture used to derive the pair creation rate (\ref{eq:paircreationrate}) only holds in the limit \(|\mu|\gg1\) which is equivalent to \(\sqrt{\gamma^2+\lambda^2}\gg\hbar\) and the assumption of relativistic particles only holds for small masses, the limit in which a comparison is possible is the strong field limit.\\
Using the pair creation rate (\ref{eq:paircreationrate}) to compute the estimated current (\ref{eq:approx_current}) in the limit \(|\mu|\gg1,\,\lambda\gg\hbar\) we find
\begin{align}
 J\approx\frac{e H}{\pi}|\mu|\exp\left[-2\pi\left(|\mu|-\frac{|\lambda|}{\hbar}\right)\right]
\end{align}
which agrees with the regularized current (\ref{eq:regcurrent}) in this limit.\\
Expanding (\ref{eq:regcurrent}) for small electric fields \(\lambda\ll\hbar\) we find
\begin{align}
 J^x_\text{reg}=\frac{eH}{\hbar^2} \frac{2\gamma \lambda}{\sinh\left(\frac{2\pi}{\hbar}\gamma\right)}+\O\left(\frac{\lambda}{\hbar}\right)^3.
\end{align}
As also visible in Fig.~\ref{fig:current} for small electric fields the mass begins to have an effect on the current. We find that for small electric fields the current gets strictly decreased by increasing mass. This is in contrast to what was found in the bosonic case of \cite{Froeb2014} and we will discuss this in the next section.
%%%%%%%%%%%%%%%%%%%%%%%%%%%%%%%%%%%%%%%%%%%%%%%%%%%%%%%%%%%%%%%%%%%%%%%%%%%%%%%%%%%%%%%%%%%%%%%%%%%%%%%%%%%%%%%%%%%%%%%%%%%%%%%%%%%%
\subsubsection{Comparison to the bosonic case}
\begin{figure}
 \centering
 \includegraphics[width=0.7\textwidth]{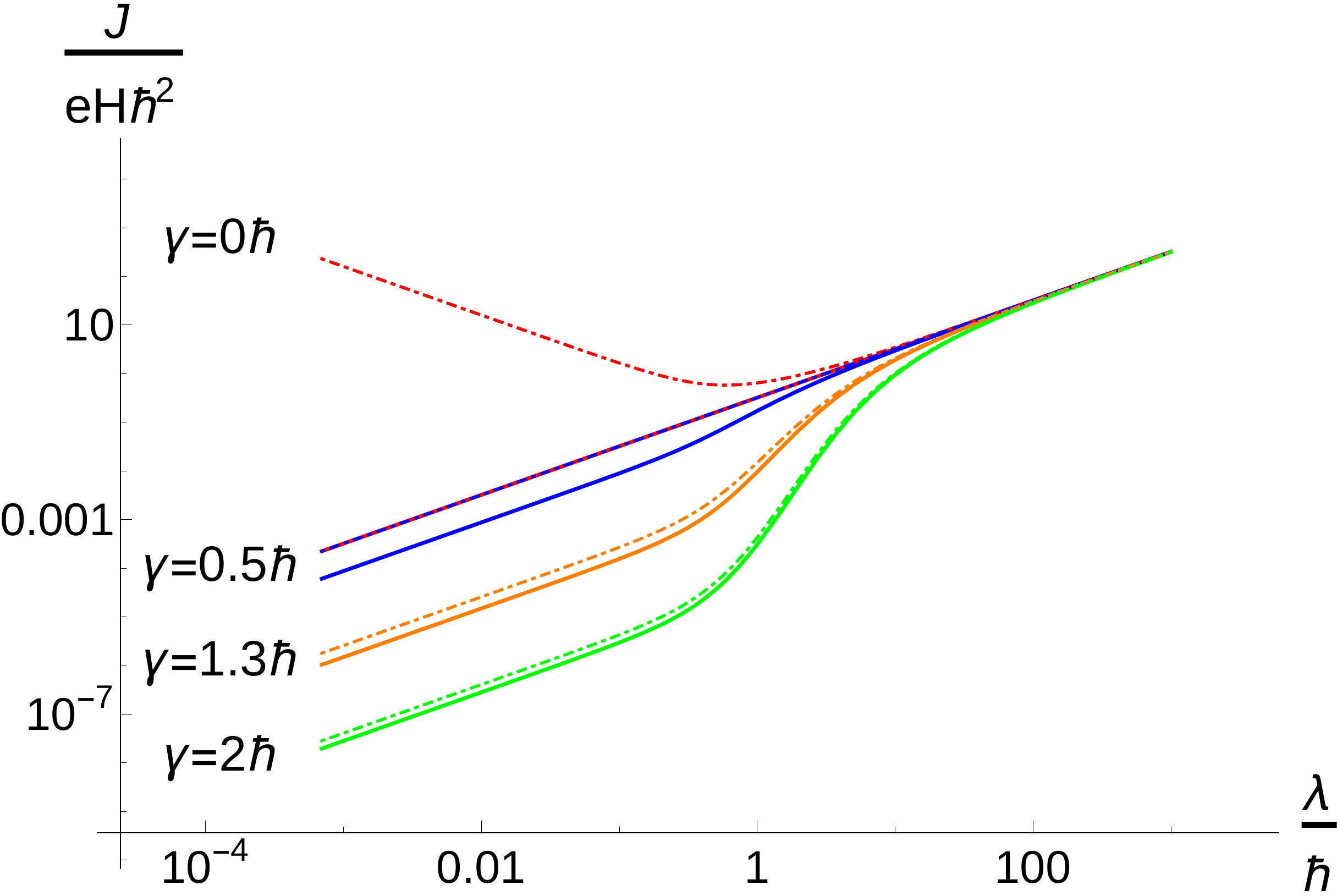} 
 % CurrentRoot.pdf: 700x426 pixel, 72dpi, 24.69x15.03 cm, bb=0 0 700 426
 \caption{Regularized current for fermions and bosons (dotted) given by (\ref{eq:regcurrent}) and (\ref{eq:bosoncurrent}) respectively, as a function of \(\lambda\) for different values of \(\gamma\). One sees that for small electric fields there is a difference while all curves have the same asymptotic limit (\ref{eq:biglambda}) for negligible mass. Observe that only the bosonic case shows the ``infrared hyper-conductivity'', i.e.~a large current for small electric field and mass. The curve with the linear response is given by \(\gamma=0\) in the fermionic case while in the bosonic case it is found to be at \(\gamma=0.5\hbar\).}
 \label{fig:current}
\end{figure}

%%%%%%%%%%%%%%%%%%%%%%%%%%%%%%%%%%%%%%%%%%%%%%%%%%%%%%%%%%%%%%%%%%%%%%%%%%%%%%%%%%%%%%%%%%%%%%%%%%%%%%%%%%%%%%%%%%%%%%%%%%%%%%%%%%%%
 \label{sec:comparison}
 We can compare the regularized current (\ref{eq:regcurrent}) to the scalar one in \(\text{dS}_2\) given by \cite{Froeb2014}
\begin{align}
 J^x_\text{boson,reg}=&\frac{eH}{\pi}\sigma\frac{\sinh\left(\frac{2\pi}{\hbar}\lambda\right)}{\sin\left(2\pi\sigma\right)}. \label{eq:bosoncurrent}
\end{align}
where \(\sigma=\sqrt{\mu^2+\frac{1}{4}}\). \\
Observe that this is the same as the fermionic current (\ref{eq:regcurrent}) if we replace \(\sigma\rightarrow\mu\).
This means that the two currents agree in the limit \(|\mu|\gg1\) since there \(\sigma\approx\mu\) (see also Fig.~\ref{fig:current}), as we have already shown for the pair creation rate in section \ref{sec:currentfromnk}. However we find a different behavior for electric fields small with respect to the Hubble constant. This is due to the fact that \(\mu\ne\sigma\) and is in difference to the flat space case where the pair creation of a constant electric field is identical for scalar and spinor QED. \\
The most striking result of the difference is, that the bosonic current is enhanced for small electric fields and values of the parameter \(\gamma<1/2\). In \cite{Froeb2014} this effect was given the name ``infrared hyper-conductivity''. We do not find this behavior for fermions. Mathematically this is due to the fact that \(\sigma\) in difference to the parameter \(\mu\) can become real for \(\gamma<1/2\). In the same way the linear response for the bosonic case is found for \(\gamma=1/2\) whereas it is found for \(\gamma=0\) for fermions. Since in flat space-time the linear response arises for massless particles the fermionic result is not as peculiar as the bosonic one.\\
The difference between \(\sigma\) and \(\mu\) comes from the last term proportional to \(\hbar^2\) in (\ref{eq:DecoupledDirac1})-(\ref{eq:DecoupledDirac2}). For the constant electric field (\ref{eq:constantfield}) it evaluates to \(1/(4\eta^2)\) and accounts for the \(1/4\) term in the Whittaker equations (\ref{eq:Whittaker1})-(\ref{eq:Whittaker2}). These terms are absent in the Klein Gordon equation and thus an additional factor of \(1/4\) has to be introduced in the variable \(\sigma\).
%%%%%%%%%%%%%%%%%%%%%%%%%%%%%%%%%%%%%%%%%%%%%%%%%%%%%%%%%%%%%%%%%%%%%%%%%%%%%%%%%%%%%%%%%%%%%%%%%%%%%%%%%%%%%%%%%%%%%%%%%%%%%%%%%%%%
\subsubsection{Vanishing gravitational field: flat spacetime limit}
%%%%%%%%%%%%%%%%%%%%%%%%%%%%%%%%%%%%%%%%%%%%%%%%%%%%%%%%%%%%%%%%%%%%%%%%%%%%%%%%%%%%%%%%%%%%%%%%%%%%%%%%%%%%%%%%%%%%%%%%%%%%%%%%%%%%
  \label{sec:wog}
  For a comparison with the flat spacetime case the parameters \(\lambda\) and \(\gamma\) defined in (\ref{eq:parameters}) are not convenient since they diverge in the flat space limit \(H\rightarrow0\). We therefore define the electric field \(\epsilon\) in units of the critical electric field as
  \begin{align}
   \epsilon:=\hbar\frac{eE}{m^2}=\hbar\frac{\lambda}{\gamma^2}. %&& \tilde{H}:=\hbar\frac{H}{m}=\frac{\hbar}{\gamma}
  \end{align}
  Performing the the flat spacetime limit \(H\rightarrow0\) of (\ref{eq:regcurrent}) we find
    \begin{align}
 \lim_{H\rightarrow0} J^x_\text{reg}&= \lim_{H\rightarrow0} \frac{e m^2}{\pi \hbar^2}\frac{\epsilon}{H} e^{-\frac{\pi}{|\epsilon|}}.
\end{align} 
 This current diverges, as it is expected for the current of an electric field in Minkowski spacetime, which was turned on at past infinity. This is due to missing Hubble dilution. One can however compare it to a current of an electric field which was turned on at finite time \(t\), as it was done in the bosonic case \cite{Kobayashi2014}. Using the substitution \(1/H\rightarrow t\) introduced above, one finds agreement with the current induced by the Schwinger effect in flat Minkowski \footnote{See e.g.~Eq.~(5.22) of \cite{Anderson2014} or Eq.~(2.14) of \cite{Anderson2014B} for the bosonic current in four dimensional flat spacetime and Eq.~(5.32) of \cite{Kluger1998} for the bosonic current in two dimensional flat spacetime.}.  
%%%%%%%%%%%%%%%%%%%%%%%%%%%%%%%%%%%%%%%%%%%%%%%%%%%%%%%%%%%%%%%%%%%%%%%%%%%%%%%%%%%%%%%%%%%%%%%%%%%%%%%%%%%%%%%%%%%%%%%%%%%%%%%%%%%%
\section{Conclusion and perspectives}
%%%%%%%%%%%%%%%%%%%%%%%%%%%%%%%%%%%%%%%%%%%%%%%%%%%%%%%%%%%%%%%%%%%%%%%%%%%%%%%%%%%%%%%%%%%%%%%%%%%%%%%%%%%%%%%%%%%%%%%%%%%%%%%%%%%%
\label{sec:conclusions}
Recently many generalizations of the original Schwinger effect  in de Sitter spacetime were proposed. While most of these studies focus on bosonic pair creation, we treat fermionic pair creation in $\text{dS}_2$ in this article. After reviewing the basic equations in section \ref{sec:preliminaries}, we perform a first derivation of the number of produced pairs based on a semiclassical saddle-point approximation. This derivation was already proposed in the more general case of $\text{dS}_4$ in \cite{us} and is straightforwardly adapted to $\text{dS}_2$. Observe that the results in section \ref{sec:SC} are independent of the specific form of the field and can be used to compute the pair creation rate for general time dependent fields.\\ 
In the rest of the paper we concentrate on the effect of a constant electric field. First we use the techniques of section \ref{sec:SC} to compute the semiclassical number of pairs per momentum mode \(k\). 
We then construct the positive and negative solutions in asymptotic past and future to use the method of Bogoliubov coefficients to compute the number of created pairs again. Here we find a first difference to the bosonic case of \cite{Froeb2014} in equation (\ref{eq:nk}). However particle creation is only well defined for well defined particle states in the asymptotic future. We show that it is the case when $|\mu| \gg 1$. In this limit the fermionic and bosonic number of created pairs agree with each other as well as with the semiclassical result.\\
This leads us for our first major conclusion: bosonic and fermionic pair creation are the same in the semiclassical approximation. In flat spacetime, this result is well known for one component fields with one pair of turning points but has been shown to be false for non constant fields with more than one component \cite{Strobel2015}. It could be interesting to explore more complicated electromagnetic field configurations in curved spacetime and check if the equivalence of semiclassical bosonic and fermionic pair creation rates still holds true.
\\
Computing the pair creation rate from the number of pairs created per momentum mode \(k\), a diverging momentum integral appears and to avoid the pair creation rate to diverge, we proposed to transform the momentum integral into a conformal time integral via a discussion of the non-adiabaticity. This heuristic step is necessary to find a coherent result, but is a sign that the pair creation rate per unit two volume is not the best quantity to describe the Schwinger effect in curved spacetime. This is due to the fact that the very definition of pair creation time does not exist.  \\
The quantity which accounts for Schwinger effect beyond the particle picture is the induced fermionic current. It is a more relevant quantity than the usual pair creation rate per unit two volume because it is not plagued by the absence of a clear definition of pair creation time. The final result is presented in (\ref{eq:regcurrent}). It has been renormalized by a adiabatic subtraction. The correction turns out to be the same as the one found in the bosonic case of \cite{Froeb2014} where the Pauli-Villars method was used.
We find that the fermionic regularized current is very similar to the one found in the bosonic case. However the effect of ``infrared hyper-conductivity'' is not present for fermions. \\
It would be interesting to investigate the relations between bosons and fermions in 4 dimensions. 
However we did not find an analytic solution of the coupled differential equation arising in this framework.  Performing a study analog to the one in this work or the bosonic case of \cite{Kobayashi2014} seems still possible numerically and is left for future work. The pair creation rate derived in the semiclassical saddle-point approximation has already been calculated and shown to agree with the bosonic one in \cite{us}.\\
As the gravitational and electrical field were taken to be external, an interesting extension would be to take their variation into account, i.e.~use the induced fermionic current in the generalized Maxwell equation and the resulting number of pairs in the Einstein equation. This could be used to find specific forms of electric fields or specific classes of spacetimes which favor or disfavor pair creation. The fact that Pauli-Villars (in the bosonic case \cite{Froeb2014}) and adiabatic regularization give rise to the same counter term, shows that pair creation is a possible framework to compare regularization techniques in curved spacetime. Other interesting issues not treated in this paper but worthy to investigate are cosmological applications of our result. The created pairs might account for the asymmetry of matter/anti-matter in our universe with a modification of an Affleck-Dine mechanism or a specific model of preheating. They could also give hints on the evolution of an accelerated period of expansion and how matter and gravitation interact in such periods.

%%%%%%%%%%%%%%%%%%%%%%%%%%%
\section*{Acknowledgements} 
The authors thank  Carlos Argüelles for many fruitful discussions. ES appreciates discussions on the issue of QFT in curved spacetimes with Alexander Stottmeister. CS and ES are supported by the Erasmus Mundus Joint Doctorate Program by Grant Number 2013-1471 and 2012-1710 from the EACEA of the European Commission respectively.
%%%%%%%%%%%%%%%%%%%%%%%%%%%

\appendix 
%%%%%%%%%%%%%%%%%%%%%%%%%%%%%%%%%%%%%%%%%%%%%%%%%%%%%%%%%%%%%%%%%%%%%%%%%%%%%%%%%%%%%%%%%%%%%%%%%%%%%%%%%%%%%%%%%%%%%%%%%%%%%%%%%%%%
\section{Properties of the Whittaker functions}
%%%%%%%%%%%%%%%%%%%%%%%%%%%%%%%%%%%%%%%%%%%%%%%%%%%%%%%%%%%%%%%%%%%%%%%%%%%%%%%%%%%%%%%%%%%%%%%%%%%%%%%%%%%%%%%%%%%%%%%%%%%%%%%%%%%%
\label{sec:Whit}
For the construction of the mode functions we use the Whittaker functions \(W_{\kappa,\mu}(z)\) and \(M_{\kappa,\mu}(z)\). These well studied functions have a number of properties which can be found in the literature (see e.g. \cite{Olver2010}). The ones we use throughout the paper are summarized in this appendix.\\
The functions have the limiting behavior for \(|z|\rightarrow\infty\)
\begin{align}
\lim_{|z|\rightarrow\infty}W_{\kappa,\mu}(z)=\lim_{|z|\rightarrow\infty}\e^{-\frac{z}{2}}z^\kappa,\hspace{1cm}\text{ for } |\text{arg(z)}|<\frac{3}{2}\pi, \label{eq:Wlimit}
\end{align}
as well as for \(z\rightarrow0\)
\begin{align}
\lim_{z\rightarrow0}M_{\kappa,\mu}(z)=\lim_{z\rightarrow0}z^{\mu+1/2}. \label{eq:Mlimit}
\end{align}
Under conjugation they behave as
\begin{align}
 \left[W_{\kappa,\mu}(z)\right]^*=W_{\kappa^*,\mu^*}(z^*),&& \left[M_{\kappa,\mu}(z)\right]^*=M_{\kappa^*,\mu^*}(z^*).\label{eq:Wconjugated}
\end{align} 
and they have the properties
\begin{align}
 W_{\kappa,-\mu}(z)=W_{\kappa,\mu}(z)&& M_{\kappa,\mu}\left(\e^{\pm\pi\ii}z\right)=\pm\ii\e^{\pm\pi\ii}M_{-\kappa,\mu}(z). \label{eq:Mproperty}
\end{align}
The function \(W_{\kappa,\mu}(z)\) can be written as an integral in the Mellin-Barnes form
\begin{align}
W_{\kappa,\mu}\left(z\right)=\e^{-\frac{z}{2}}\int_{-\ii\infty}^{\ii\infty}\frac{ds}{2\ii\pi}\frac{\Gamma\left(\frac12+\mu+s\right)\Gamma\left(\frac12-\mu+s\right)\Gamma\left(-\kappa-s\right)}{\Gamma\left(\frac12+\mu-\kappa\right)\Gamma\left(\frac12-\mu-\kappa\right)}z^{-s}. \label{eq:MellinBarnes}
\end{align}
The two functions are connected through
\begin{align}
W_{\kappa,\mu}(z)=\frac{\Gamma(-2\mu)}{\Gamma(\frac{1}{2}-\mu-\kappa)}M_{\kappa,\mu}(z)+\frac{\Gamma(2\mu)}{\Gamma(\frac{1}{2}+\mu-\kappa)}M_{\kappa,-\mu}(z) \label{eq:WtoM}.
\end{align}
The Wronskian is given by
\begin{align}
W_{\kappa,\mu}(z)\frac{dW_{-\kappa,-\mu}\left(\e^{\pm\ii\pi}z\right)}{dz}-\frac{dW_{\kappa,\mu}(z)}{dz}W_{-\kappa,-\mu}\left(\e^{\pm\ii\pi}z\right)=\e^{\mp\ii\pi\kappa}. \label{eq:Wronskian}
\end{align} 
By using \(\Gamma(n+1)=\Gamma(n)n\) in (\ref{eq:MellinBarnes}) one can find
\begin{align}
 W_{\mu,\kappa-\frac{1}{2}}(z)=\frac{2\kappa+1-z}{2(\mu^2-\kappa^2)} W_{\mu,\kappa+\frac{1}{2}}(z)-\frac{z}{(\mu^2-\kappa^2)}\frac{dW_{\mu,\kappa+\frac{1}{2}}(z)}{dz}. \label{eq:pmonehalfIdentity}
\end{align}
%%%%%%%%%%%%%%%%%%%%%%%%%%%%%%%%%%%%%%%%%%%%%%%%%%%%%%%%%%%%%%%%%%%%%%%%%%%%%%%%%%%%%%%%%%%%%%%%%%%%%%%%%%%%%%%%%%%%%%%%%%%%%%%%%%%%
\section{Wronskian condition}\label{sec:WronskianCondition}
%%%%%%%%%%%%%%%%%%%%%%%%%%%%%%%%%%%%%%%%%%%%%%%%%%%%%%%%%%%%%%%%%%%%%%%%%%%%%%%%%%%%%%%%%%%%%%%%%%%%%%%%%%%%%%%%%%%%%%%%%%%%%%%%%%%%
In this appendix we show how to derive the value of parameters (\ref{eq:C_14}) of the positive and negative frequency solutions at asymptotic past (\ref{eq:2DFermSol}) so that the Wronkskian condition (\ref{eq:WronskianCondition}) holds. We therefore define 
\begin{align}
M:=\psi^+(\eta)\psi^+(\eta)^\dagger+\psi^-(\eta)\psi^-(\eta)^\dagger.%=\frac{\hbar}{a(\eta)}\mathbbm{1}.\label{eq:WronskianCondition}
\end{align}
Now we can use the behavior of the Whittaker function under conjugation given by (\ref{eq:Wconjugated}) and the specific form of the solutions (\ref{eq:2DFermSol}) given in (\ref{eq:Csolutions}) and (\ref{eq:C_23}) to find
\begin{align}
 \psi_1^+(\eta)^*=\psi_2^-(\eta)\cdot\begin{cases}
                  -\frac{C_2^*}{C_3} &\text{ for } k>0\\
                  +\frac{C_3^*}{C_2} &\text{ for } k<0
                 \end{cases},&&
 \psi_1^-(\eta)^*=\psi_2^+(\eta)\cdot\begin{cases}
                  +\frac{C_3^*}{C_2} &\text{ for } k>0\\
                  -\frac{C_2^*}{C_3} &\text{ for } k<0
                 \end{cases},  \label{eq:psiconjugated}
\end{align}
where \(\psi_1^\pm(\eta)\) and \(\psi_2^\pm(\eta)\) are the first and second component of \(\psi_\text{in}^\pm\) respectively. 
This can be used to find
\begin{align}
 {M_{12}}^*=M_{21}%=&\psi_1^+(\eta)^*\psi_2^+(\eta)+\psi_1^-(\eta)^*\psi_2^-(\eta)\\
 =&\psi_2^+(\eta)\psi_2^-(\eta)\left[\frac{C_3^*}{C_2}-\frac{C_2^*}{C_3}\right].
\end{align}
Now requiring the Wronskian condition (\ref{eq:WronskianCondition}), i.e. \(M_{12}=M_{21}=0\), we find
\begin{align}
 |C_2|^2=|C_3|^2. \label{eq:C_14^2}
\end{align}
Using (\ref{eq:psiconjugated}) for the diagonal elements of \(M\) we find
\begin{align}
 M_{11}={M_{22}}^*=-\frac{C_3^*}{C_2}\sgn(k)\left[\psi_1^+(\eta)\psi_2^-(\eta)-\psi_1^-(\eta)\psi_2^+(\eta)\right].
\end{align}
We can now use the Dirac equation (\ref{eq:CoupledDirac1}) to bring this into the form
\begin{align}
 M_{11}={M_{22}}^*=-\frac{\ii\hbar}{m a(\eta)}\frac{C_3^*}{C_2}\sgn(k)\left[\psi_1^+(\eta){\psi_1^-}'(\eta)-\psi_1^-(\eta){\psi_1^+}'(\eta)\right].
\end{align}
Using the solutions (\ref{eq:2DFermSol}) we find 
\begin{align}
 M_{11}={M_{22}}^*=-\frac{\ii\hbar}{m a(\eta)}\frac{C_3^*}{C_2}\left[\psi_1^a(z)\frac{d}{d\eta}\psi_1^b(z)-\psi_1^b(z)\frac{d}{d\eta}\psi_1^a(z)\right],\end{align}
using the explicit form of the solutions (\ref{eq:Csolutions}) and (\ref{eq:C_23}) this becomes
\begin{align}
 M_{11}={M_{22}}^*=-\ii\frac{|C_3|^2}{ a(\eta)}\frac{2k}{H\hbar}\left[W_{\kappa-\frac{1}{2},\mu}(z)\frac{dW_{-\kappa+\frac{1}{2},-\mu}(-z)}{dz}-\frac{dW_{\kappa-\frac{1}{2},\mu}(z)}{dz}W_{-\kappa+\frac{1}{2},-\mu}(-z)\right].
 \end{align}
 Using the Wronskian (\ref{eq:Wronskian}) we find
 \begin{align}
 M_{11}={M_{22}}^*=-\ii\frac{|C_3|^2}{ a(\eta)}\frac{2k}{H\hbar}\e^{-i\pi\sgn(k)\left(\kappa-\frac{1}{2}\right)}=\frac{|C_3|^2}{ a(\eta)}\frac{2|k|}{H\hbar}\e^{-i\pi\sgn(k)\kappa}.
 \end{align}
Now requiring the Wronskian condition (\ref{eq:WronskianCondition}), i.e.~\(M_{11}=M_{22}=a(\eta)^{-1}\), we find
\begin{align}
 |C_3|^2=\frac{\hbar H}{2|k|}\e^{i\pi\kappa\sgn(k)}.
\end{align}
Choosing a physically irrelevant phase this leads to (\ref{eq:C_14}) using (\ref{eq:C_14^2}). \\
Observe that from (\ref{eq:psiconjugated}) and (\ref{eq:C_14^2}) we find
\begin{align}
 |\psi_1^+(\eta)|^2=|\psi_2^-(\eta)|^2,&&|\psi_2^+(\eta)|^2=|\psi_1^-(\eta)|^2. \label{eq:psi+psi-}
\end{align}

%%%%%%%%%%%%%%%%%%%%%%%%%%%%%%%%%%%%%%%%%%%%%%%%%%%%%%%%%%%%%%%%%%%%%%%%%%%%%%%%%%%%%%%%%%%%%%%%%%%%%%%%%%%%%%%%%%%%%%%%%%%%%%%%%%%%
\section{Computation of the integral for the current}
%%%%%%%%%%%%%%%%%%%%%%%%%%%%%%%%%%%%%%%%%%%%%%%%%%%%%%%%%%%%%%%%%%%%%%%%%%%%%%%%%%%%%%%%%%%%%%%%%%%%%%%%%%%%%%%%%%%%%%%%%%%%%%%%%%%%
\label{sec:Integral}
If we change the variable in (\ref{eq:current2}) to \(v:=|k|\eta/\hbar\) we find
\begin{align}
 J^x&=-\frac{e }{\eta}\int_0^{\infty}\frac{dv}{2\pi}\Big(|\psi_1^+(\eta)|^2\Big|_{k>0}-|\psi_2^+(\eta)|^2\Big|_{k>0}+|\psi_1^+(\eta)|^2\Big|_{k<0}-|\psi_2^+(\eta)|^2\Big|_{k<0}\Big),\\
     &=-\frac{2 e }{\eta}\int_0^{\infty}\frac{dv}{2\pi}\Big(|\psi_1^+(\eta)|^2\Big|_{k>0}-|\psi_2^+(\eta)|^2\Big|_{k<0}\Big),
\end{align}
where we used the normalization (\ref{eq:normalization}) of the modes. Using the specific form of the solutions (\ref{eq:2DFermSol2}) this takes the form
\begin{align}
 J^x&=-\frac{\gamma^2}{\hbar^2}\frac{He}{2\pi}\int_0^\infty\frac{dv}{v}\sum_{r=-1,1}r\,\e^{-r\pi\ii\kappa}\left|W_{ r \kappa-\frac12,r\mu}(2\ii v)\right|^2.
\end{align}
Using the Mellin-Barnes form of the Whittaker function (\ref{eq:MellinBarnes}) we can write the current as
\begin{align}
 \begin{split}
 J^x=-&\frac{\gamma^2}{\hbar^2}\frac{He}{2\pi}\lim_{\xi\rightarrow\infty}\int_0^\xi\frac{dv}{v}\int_{-\ii\infty}^{\ii\infty}\frac{ds}{2\ii\pi}\int_{-\ii\infty}^{\ii\infty}\frac{dt}{2\ii\pi}(2\ii v)^{-s}(-2\ii v)^{-t}\sum_{r=-1,1}r\,e^{-r\pi\ii \kappa}\\
 &\times\frac{\Gamma(1/2+\mu+s)\Gamma(1/2-\mu+s)\Gamma(1/2-r\kappa-s)}{\Gamma(1+\mu-r\kappa)\Gamma(1-\mu-r\kappa)}
 \frac{\Gamma(1/2-\mu+t)\Gamma(1/2+\mu+t)\Gamma(1/2+r\kappa-t)}{\Gamma(1-\mu+r\kappa)\Gamma(1+\mu+r\kappa)}.
 \end{split}
 \end{align}
 We now perform the integral over \(v\) 
 \begin{align}
 \begin{split}
 J^x=&-\frac{\gamma^2}{\hbar^2}\frac{He}{(2\pi)^3}\lim_{\xi\rightarrow\infty}\frac{1}{\Gamma(1+\mu-\kappa)\Gamma(1-\mu-\kappa)\Gamma(1-\mu+\kappa)\Gamma(1+\mu+\kappa)}\\
 &\times\int_{-\ii\infty}^{\ii\infty}{ds}\,\sum_{r=-1,1}r\,e^{-r\pi\ii\kappa}e^{-\ii\frac{\pi}{2}s}\,\Gamma(1/2+\mu+s)\Gamma(1/2-\mu+s)\Gamma(1/2-r\kappa-s)\\
 &\times\int_{-\ii\infty}^{\ii\infty}{dt}\,\e^{\ii\frac{\pi}{2}t}\,{\Gamma(1/2-\mu+t)\Gamma(1/2+\mu+t)\Gamma(1/2+r\kappa-t)}\frac{(2\xi)^{-s-t}}{s+t}
 \end{split}
 \end{align}
We can close the contour of the integral over \(t\) in the \(t>0\) -plane in order to use the residue theorem. The contributing poles thus are \(t=-s\) and \(t=-r\kappa+1/2+n\).  Due to the \(\xi\rightarrow\infty\) limit the integral is only non-zero for \(t+s\le0\). Since we could close the integral over \(s\) in a similar way in the \(s>0\) -plane for \(t\ne-s\) the only pole which gives non-zero contribution is \(t=-s\). Using the residue theorem we thus find
 \begin{align}
 \begin{split}
 J^x=-&\frac{\gamma^2}{\hbar^2}\frac{\ii eH}{(2\pi)^2}\int_{-\ii\infty}^{\ii\infty}{ds}\,\,\frac{\Gamma(1/2+\mu+s)\Gamma(1/2-\mu+s)
 \Gamma(1/2-\mu-s)\Gamma(1/2+\mu-s)}{\Gamma(1+\mu-\kappa)\Gamma(1-\mu-\kappa)\Gamma(1-\mu+\kappa)\Gamma(1+\mu+\kappa)}\\
 &\hspace{3cm}\times\sum_{r=-1,1}r\,\e^{-\ii{\pi}(r\kappa+s)}\Gamma(1/2-r\kappa-s)\Gamma(1/2+r\kappa+s)
 \end{split},\\
 =&-\frac{\gamma^2}{\hbar^2}\frac{1}{\mu^2-\kappa^2}\frac{eH}{4\pi}\ii\int_{-\ii\infty}^{\ii\infty}{ds}\,\frac{\sin[(\mu+\kappa)\pi]\sin[(\mu-\kappa)\pi]}{\cos[(\mu+s)\pi]\cos[(\mu-s)\pi]}\sum_{r=-1,1}r\,\frac{\e^{-\ii{\pi}(r\kappa+s)}}{\cos[(r\kappa+s)\pi]}.
 \end{align}
 To solve this integral we now write trigonometric functions as exponentials and change the integration variable to \(X:=\exp(\ii\pi s)\). This leads to a integral which can be solved using the standard decomposition theorems for rational fractions
 \begin{align}
 J^x=-&\frac{\gamma^2}{\hbar^2}\frac{4}{\mu^2-\kappa^2}\frac{eH}{\pi^2}\ii\int_{\infty}^{0}{dX}\,\frac{\sin[2\pi\kappa]\sin[(\mu+\kappa)\pi]\sin[(\mu-\kappa)\pi]X^4}{(X^2+\e^{2\ii\pi\kappa})(X^2+\e^{-2\ii\pi\kappa})(X^2+\e^{2\ii\pi\mu})(X^2+\e^{-2\ii\pi\mu})},\\
 =&\frac{eH}{\pi}\ii\left(\mu\frac{\sin(2\pi\kappa)}{\sin(2\pi\mu)}-\kappa\right).
 \end{align}

%\bibliography{DeSitter}

\end{document}